\documentclass[prd,tightenlines,nofootinbib,superscriptaddress]{revtex4}

\usepackage{amsfonts,amssymb,amsthm,bbm}

\usepackage{amsmath}

\usepackage{hyperref}

\usepackage{booktabs} 
\usepackage{array}

\usepackage{color,psfrag}
\usepackage[dvips]{graphicx}
\usepackage{tikz}
\usetikzlibrary{calc}
\usetikzlibrary{decorations.pathmorphing}
\usetikzlibrary{shapes.geometric}
\usetikzlibrary{arrows,decorations.markings}
\usepackage[abs]{overpic}

\usepackage[caption=false]{subfig}

\newcolumntype{M}[1]{>{\centering\arraybackslash}m{#1}}

\newcommand{\blue}{\color{blue}}
\newcommand{\purple}{\color{purple}}

\newcommand{\orange}{\color{orange}}

\newcommand{\C}{{\mathbb C}}

\newcommand{\R}{{\mathbb R}}

\newcommand{\cG}{{\mathcal G}}

\newcommand{\cT}{{\mathcal T}}

\newcommand{\cD}{{\mathcal D}}

\newcommand{\cS}{{\mathcal S}}

\newcommand{\cY}{{\mathcal Y}}

\newcommand{\be}{\begin{equation}}
\newcommand{\ee}{\end{equation}}
\newcommand{\beq}{\begin{eqnarray}}
\newcommand{\eeq}{\end{eqnarray}}
\newcommand{\bes}{\begin{eqnarray}}
\newcommand{\ees}{\end{eqnarray}}

\newcommand{\f}{\frac}

\def\nn{\nonumber}

\newcommand{\w}{\wedge}

\def\vphi{\varphi}

\def\tvphi{\tilde{\vphi}}
\def\ttheta{\tilde{\theta}}
\def\tY{\widetilde{Y}}

\newcommand{\bllink}[2]
{\draw[orange, decoration={markings,mark=at position 0.6 with {\arrow[scale=1.5,>=stealth]{>}}},postaction={decorate}] (#1) --(#2)}

\def\centerarc[#1](#2)(#3:#4:#5)
{ \draw[#1] ($(#2)+({#5*cos(#3)},{#5*sin(#3)})$) arc (#3:#4:#5); }

\def\centerarcnodes[#1](#2)(#3:#4:#5)(#6,#7)
{\coordinate(#6) at ($(#2)+({#5*cos(#3)},{#5*sin(#3)})$);
	\coordinate(#7) at ($(#2)+({#5*cos(#4)},{#5*sin(#4)})$);
	\draw[#1] ($(#2)+({#5*cos(#3)},{#5*sin(#3)})$) arc (#3:#4:#5); }

\def\angcircle(#1)(#2)(#3:#4) {\coordinate(#1) at ($(#2)+({#4*cos(#3)},{#4*sin(#3)})$); }

\begin{document}

\title{Geometric Formula for 2d Ising Zeros: Examples \& Numerics}

\author{{\bf I\~naki Garay}}\email{inaki.garay@ehu.eus}
\affiliation{Department of Physics and EHU Quantum Center, University of the Basque Country UPV/EHU, Barrio Sarriena s/n, 48940, Leioa, Spain}

\author{{\bf Etera R. Livine}}\email{etera.livine@ens-lyon.fr}
\affiliation{Univ Lyon, Ens de Lyon, Universit\'e Claude Bernard, CNRS,
Laboratoire de Physique, F-69342 Lyon, France}

\date{\today}

\begin{abstract}

A geometric formula for the zeros of the partition function of the  inhomogeneous 2d Ising model was recently proposed in terms of the angles of 2d triangulations embedded in the flat 3d space. Here we proceed to an analytical check of this formula on the cubic graph, dual to a double pyramid, and provide a thorough numerical check  by generating random 2d planar triangulations.
Our method is to generate Delaunay triangulations of the 2-sphere then performing random local rescalings. For every 2d triangulations, we compute the corresponding Ising couplings from the triangle angles and the dihedral angles, and check directly that the Ising partition function vanishes for these couplings (and grows in modulus in their neighborhood). In particular, we lift an ambiguity of the original formula on the sign of the dihedral angles and establish a convention in terms of convexity/concavity.
Finally, we extend our numerical analysis to 2d toroidal triangulations and show that the geometric formula does not work and will need to be generalized, as originally expected, in order to accommodate for non-trivial topologies.

\end{abstract}

\maketitle
\tableofcontents

\section*{Introduction}

Over the past few years, theoretical physics has developed a very fruitful interface between the thriving field of condensed matter and the burgeoning line of research of quantum gravity, sharing methods of statistical physics and quantum information, to reveal insights for instance into quantum field correlations, topological phases, multi-bodied entanglement, the physics of edge modes and holographic dualities.

One of those intriguing outcomes is an exact and explicit duality between 3d quantum gravity (QG) and the 2d Ising model \cite{Bonzom:2015ova}. Actually, several such dualities between bulk gravity theories and boundary gauge field theories have been uncovered in the context of the search for a holographic formulation of classical and quantum gravity. Most of those holographic dualities are usually formulated in terms of the path integral amplitudes of continuum field theories (see e.g. \cite{Castro:2011zq} for dualities between 3d quantum gravities and 2d conformal field theories). What distinguishes the QG${}_{3d}$$\,\leftrightarrow\,$Ising${}_{2d}$ BCL duality presented in \cite{Bonzom:2015ova} is based on an exact equality between the fully regularized path integral amplitudes of 3d quantum gravity with boundary and the (inverse square of the) 2d inhomogeneous Ising partition function, which holds at the discrete level, even away from criticality.
This builds on an earlier result by Westbury showing that the (inverse square of the) 2d Ising partition function provides a generating function for spin network evaluations  \cite{Westbury1998,CostantinoHDR}.

The origin of this duality is that 3d  gravity is a topological field theory, which can be exactly quantized as a discrete path integral. By ``exact'', it is meant that this path integral, even though discrete, captures all the symmetries and physical degrees of freedom of 3d gravity. This yields a topological quantum field theory (TQFT), with no local degree of freedom in the bulk, which is naturally holographic:  all physical degrees of freedom are faithfully represented on the boundary. Then \cite{Bonzom:2015ova} identified a class of quantum boundary states such that the induced boundary theory is dual to a double copy of the Ising model,
Those coherent boundary states allow for an exact evaluation of the 3d quantum gravity amplitudes and admit a critical regime where they exhibit scale-invariance properties \cite{Freidel:2012ji,Bonzom:2012bn,Dittrich:2013jxa} (see also \cite{Livine:2022ptr} for more details on the structure boundary states).
Surprisingly, this duality is realized through a supersymmetry between the bosonic gravitational degrees of freedom of the geometry and the fermionic Ising spins, even though there is no supergravity to begin with.

Since the Ising model is much better understood and studied than quantum gravity, those previous works naturally open the door to using our expertise in statistical physics to explore the behavior and phase diagram of the quantum gravity path integral. Going in the other direction, from quantum gravity to statistical physics, it turned out that an interesting side-product of this duality is actually a geometric formula for complex zeros of the Ising partition function \cite{Bonzom:2024zka}. These zeros, à la Lee-Yang, for finite graph become critical couplings in the thermodynamic limit of large lattices. Let us underline that we are dealing with the inhomogeneous Ising model, with different couplings living on each link.

This geometric Ising zero formula was derived in \cite{Bonzom:2024zka} by looking at the saddle points of the quantum gravity path integral in its semi-classical regime. It expresses the zeros of the Ising partition function on a planar graph in terms of the geometry of its dual 2d triangulation embedded in the flat 3d  Euclidean space. It relies on the analysis of the asymptotics of two factors of the path integral - the measure and the amplitude. The control over those approximation unfortunately lack strict mathematical rigour and the formula remains a conjecture\footnotemark.
\footnotetext{
After the present work was first posted on the arXiv, a mathematical proof of the results appeared in \cite{Lis:2024diu}, confirming our analytical exploration and numerical analysis, presented here, and setting a mathematically-rigorous framework for the zeros' formula.
}

\medskip

The purpose of the present work is to support that claim of a geometric formula for the zeros of the 2d Ising model. Our approach is two-fold.
On the one hand, we provide examples to illustrate how the formula works, with simple graphs dual to visualisable 2d structures: the triangular pancake,  tetrahedron, the pyramid, double pyramid and cube. We can perform an analytical check of the formula in those cases.
It also allows us to test the generalization of the formula to circle patterns.
On the other hand, in order to test the formula in more general settings, we perform a numerical check on random triangulations, with the topology of a 2-sphere. These are realized in practice as Delaunay triangulations of the 2-sphere, with an extra-rescaling allowing to explore both convex and concave configurations. This provides us with a thorough validation of the geometric formula for Ising zeros.
Moreover, it lifts a sign ambiguity in the original formula on the dihedral angles and shows that the correct sign of the dihedral angle around an edge reflect the convexity or concavity of the gluing of faces at that edge.

Finally, we conclude this study with a numerical exploration of the toroidal topology. Indeed, the QG${}_{3d}$$\,\leftrightarrow\,$Ising${}_{2d}$  duality of \cite{Bonzom:2015ova} was explicitly formulated from the topology of 3-ball bounded by a 2-sphere, so that the geometric formula for the Ising zeros proposed in \cite{Bonzom:2024zka} is based on the assumption of a spherical surface topology and planar graphs. It is not clear whether or not it could be applied straightforwardly to higher genus topologies. Therefore, we test it on a prismatic torus, made of three prisms glued together into a ring with adjustable edge lengths. This shows that the geometric formula does not hold for a toroidal topology, that topology does matter and that the formula will need to be suitably generalized to non-trivial topologies beyond the sphere.

\section{Geometric Formula for Ising Zeros}

Let us study the inhomogeneous Ising model on a graph $\Gamma$:
\be
Z_{\Gamma}[\{y_{\ell}\}]
=
\sum_{\{\sigma_{n}=\pm\}}\prod_{\ell} e^{y_{\ell}\sigma_{s(\ell)}\sigma_{t(\ell)}}\,,
\label{eq_partition1}
\ee
where $\ell$ labels the links of the graph $\Gamma$ and $n$ its nodes. The $y_{\ell}$ are the Ising couplings between nearest neighbours and are arbitrary complex numbers. Finally, $s(\ell)$ and $t(\ell)$ label the \textit{source} and \textit{target} vertices defining the link $\ell$.
Using that $e^{y\sigma}=\cosh y+\sigma\sinh y$ for arbitrary $y\in\C$ and $\sigma=\pm$, we can re-write the Ising partition function as:
\be
Z_{\Gamma}[\{y_{\ell}\}]
=
\left(\prod_{\ell}\cosh y_{\ell}\right)\,
\sum_{\{\sigma_{n}=\pm\}} \prod_{\ell}\,(1+\sigma_{s(\ell)}\sigma_{t(\ell)}\tanh y_{\ell})
=
\f{2^{N_{n}}}{\prod_{\ell}
\sqrt{1-Y_{\ell}^{2}}}\,P_{\Gamma}[\{Y_{\ell}\}]\,,\label{eq:partition_looppol}
\ee
where $N_{n}$ is the number of nodes of the graph and $P_{\Gamma}$ is a polynomial in redefined couplings $Y_{\ell}=\tanh y_{\ell}$:
\be
P_{\Gamma}[\{Y_{\ell}\}]
=
\sum_{\cG_{\textrm{even}}\subset \Gamma} \,\,\prod_{\ell\in\cG}Y_{\ell}\,,
\ee
with the sum over all even subgraphs $\cG$, i.e. that the valence of every node of $\cG$ is even \cite{Bonzom:2015ova}.
This sum always starts with 1, corresponding to the trivial empty subgraph.

Here, we focus on the 2d Ising model, so we restrict ourselves to planar graphs, i.e. that can be drawn on a 2-sphere. And we will further restrict ourselves to 3-valent graphs. In that case, the sum over even subgraphs becomes a sum over all unions of (disjoint) cycles on the graph $\Gamma$, and we will refer to $P_{\Gamma}$ as the loop polynomial.

\begin{figure}[ht!]

\centering
	\begin{tikzpicture}[scale=2.5]

		\coordinate (A1) at (-0.39,0.81);
		\coordinate (A2) at (-0.55,0.19);
		\coordinate (A3) at (-0.25,-0.58);
		\coordinate (A4) at (0.4,-0.78);

		\coordinate (A9) at (1.26, 0.76);
		\coordinate (A10) at (0.64, 1.01);
		\coordinate (I1) at (0.2, 0.2);
		\coordinate (I2) at (1, -0.24);

\draw (A1) node{$\bullet$};
\draw (A2) node{$\bullet$}; 
\draw (A3) node{$\bullet$};
\draw (A4) node{$\bullet$}; 

\draw (A9) node{$\bullet$}; 
\draw (A10) node{$\bullet$}; 
\draw (I1) node{$\bullet$}; 
\draw (I2) node{$\bullet$};

		\draw(A1)--(A2);
		\draw(A2)--(A3);
		\draw(A3)--(A4);

		\draw(A9)--(A10);
		\draw(A10)--(A1);
		\draw(A1)--(I1);
		\draw(A2)--(I1);
		\draw(A3)--(I1);
		\draw(A4)--(I1); 
		\draw(A4)--(I2); 

		\draw(A9)--(I2);
		\draw(I1)--(I2);
		\draw(I2)--(A10);
		\draw(I1)--(A10);

\coordinate (B1) at (0.2, 0.7);
\coordinate (B1b) at (.63, 0.22);
\coordinate (B9) at (-0.25, 0.4); 
\coordinate (B9b) at (-0.9, 0.6); 
\coordinate (B10) at (0.03, 1.21);
\coordinate (B2) at (0.95, 0.5);
\coordinate (B6b) at (0.6, -0.3);
\coordinate (B2b) at (1.45, 0.2);
\coordinate (B11) at (1.1, 1.19);
\coordinate (B7) at (0.1, -0.35);
\coordinate (B6) at (1.1, -0.8);
\coordinate (B8) at (-0.2, -0.1);
 \coordinate (B17) at (-0.06, -1);
 \coordinate (B18) at (-0.62, -0.44);

\bllink{B1}{B1b};
\bllink{B1}{B9};
\bllink{B1}{B10};
\bllink{B1b}{B2};
\bllink{B1b}{B6b};
\bllink{B2}{B2b};
\bllink{B2}{B11};
\bllink{B6b}{B6};
\bllink{B7}{B6b};
\bllink{B7}{B8};
\bllink{B7}{B17};
\bllink{B8}{B9};
\bllink{B8}{B18};
\bllink{B9}{B9b};

\draw (0.5, 0.5) node[orange]{{ $\ell$}};
\draw (.48, .85) node{{$e$}};
		
\draw[orange] (B1) node{$\bullet$};
\draw[orange] (B1b) node{$\bullet$};
\draw[orange] (B10) node{$\bullet$};
\draw[orange] (B9) node{$\bullet$};
\draw[orange] (B9b) node{$\bullet$};
\draw[orange] (B2) node{$\bullet$};
\draw[orange] (B6b) node{$\bullet$};
\draw[orange] (B2b) node{$\bullet$};
\draw[orange] (B11) node{$\bullet$};
\draw[orange] (B6) node{$\bullet$};
\draw[orange] (B7) node{$\bullet$};
\draw[orange] (B8) node{$\bullet$};
\draw[orange] (B17) node{$\bullet$};
\draw[orange] (B18) node{$\bullet$};

	\end{tikzpicture}

\caption{
A 3-valent graph $\Gamma$ (in {\orange orange}) and its dual triangulation: graph nodes are dual to triangles, and  graph links $\ell$ are dual to triangulation edges $e$.
The orientation of the links does not matter here and the source and target vertices play the same role in the Ising model.
The triangulation is not necessarily in a single 2d plane. It can be folded and have a non-trivial embedding in the 3d space $\R^{3}$, with non-zero dihedral angles between neighbour triangles.
}
\label{fig:dualgraph}
\end{figure}
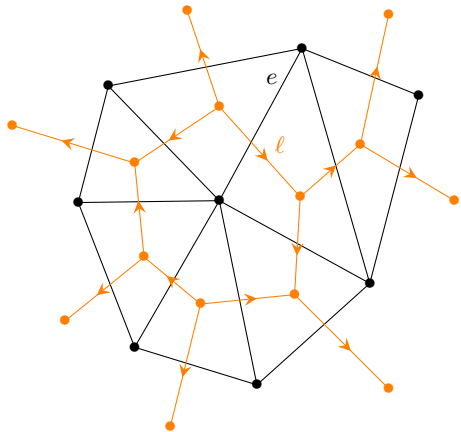

Previous work \cite{Bonzom:2015ova,Bonzom:2019dpg,Bonzom:2024zka} conjectured a formula for the zeros of the 2d Ising model, based on a saddle point analysis of 3d quantum gravity amplitudes in the semi-classical regime. The formula is constructed in terms of 2d triangulations embedded in the 3d flat space. The triangulations are dual to the graph $\Gamma$, so a triangle is dual to a graph node, an edge is dual to a graph link and a vertex is dual to a graph loop, as illustrated on fig.\ref{fig:dualgraph}.

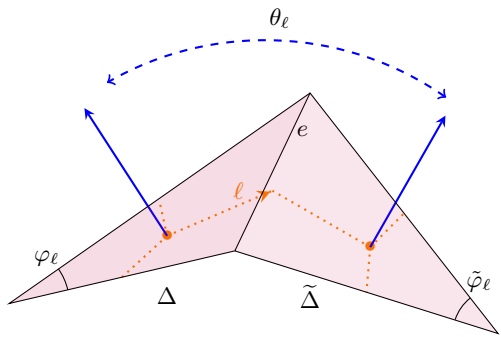
\begin{figure}[h!]

\centering
\begin{tikzpicture}[scale=1]

\coordinate(a) at (0,0.4) ;
\coordinate(b) at (1,2.5);
\coordinate(ab) at (0.5,1.2) ;
\coordinate(c1) at (3.5,-0.7);
\coordinate(c2) at (-3,-0.3);
\coordinate(ac1) at (1.75,-0.15);
\coordinate(ac2) at (-1.5,0.05);
\coordinate(bc1) at (2.25,0.9);
\coordinate(bc2) at (-1,1.05);
\coordinate(e1) at (1.8,0.45);
\coordinate(e2) at (-0.9,0.6);
\coordinate(f1) at (2.8,2.2);
\coordinate(f2) at (-2,2.3);
\coordinate(O) at (0.4,-1);

\draw[orange] (e1) node {$\bullet$};
\draw[orange] (e2) node {$\bullet$};

\draw (ab)++(-0.45,0) node[orange] {$\ell$};
\draw (b)++(-0.1,-.5) node {$e$};
\draw (a)++(-.9,-.6) node {$\Delta$};
\draw (a)++(1,-.6) node {$\widetilde{\Delta}$};
\draw (c1)++(-.25,.69) node {$\tilde{\vphi}_{\ell}$};
\draw (c2)++(.5,.6) node {$\vphi_{\ell}$};
\draw (O)++(0.2,4.5) node {$\theta_{\ell}$};

\draw[orange,thick,dotted] (e1)--(ab);
\draw[orange,thick,dotted] (e1)--(ac1);
\draw[orange,thick,dotted] (e1)--(bc1);
\draw[orange,thick,dotted, decoration={markings,mark=at position 0.98 with {\arrow[scale=1.2,>=stealth]{>}}},postaction={decorate}] (e2)--(ab);
\draw[orange,thick,dotted] (e2)--(ac2);
\draw[orange,thick,dotted] (e2)--(bc2);

\draw[blue,->, thick,>=stealth](e1)--(f1) ;
\draw[blue,->, thick,>=stealth](e2)--(f2) ;
\draw (a)--(b)  --(c1)--(a);
\draw (b)--(c2)--(a);
\fill[fill=purple,fill opacity=0.1]  (a)--(b)  --(c1)--(a);
\fill[fill=purple,fill opacity=0.14]  (a)--(b)  --(c2)--(a);

\centerarc[](c1)(128:162:.6);
\centerarc[](c2)(13:35:.8);

\centerarc[blue,thick,dashed,<->=0.5](O)(55:120:4.2);

\end{tikzpicture}

\caption{
The two adjacent triangles $\Delta$ and $\widetilde{\Delta}$ sharing the edge $e$, are dual to two graph nodes connected by the graph link $\ell$ (in dotted line).
The 2d angles $\vphi_{\ell}$ and $\tilde{\vphi}_{\ell}$ are the triangle angles at the vertices opposite to the edge.
The dihedral angle $\theta_{\ell}$ between the two triangles is the angle between their normal vectors, and reflects the non-trivial extrinsic curvature of the embedding of the 2d triangulation in the flat 3d space. 
\label{fig:angles}}

\end{figure}

More precisely, it was conjectured that couplings $Y_{\ell}$ constructed in terms of the triangulation angles are roots of the loop polynomial $P_{\Gamma}$ and thereby zeros of the Ising partition function:
\be
Y_{\ell}^{(0)}
=
e^{\pm i\f{\theta_{\ell}}2}\,
\sqrt{\tan\f{\vphi^{s}_{\ell}}2\,\tan\f{\vphi^{t}_{\ell}}2}
\quad
\overset{??}\Longrightarrow
\quad
P_{\Gamma}[\{Y_{\ell}^{(0)}\}]=0
\,,
\label{eq_Ising_zeroes}
\ee
where $\pm$ is an overall sign in front of all the dihedral angles.
As drawn on fig.\ref{fig:angles}, the angles are all defined on the triangulations around the edge $e$ dual to the considered link $\ell$:
$\theta_{\ell}$ is the dihedral angle between the two vectors  normal  to the triangles, while $\vphi^{s}_{\ell}$ and $\vphi^{t}_{\ell}$ are the triangle angles at the vertices opposite to the edge $e$.

\begin{figure}[ht!]
\centering
\includegraphics[height=40mm]{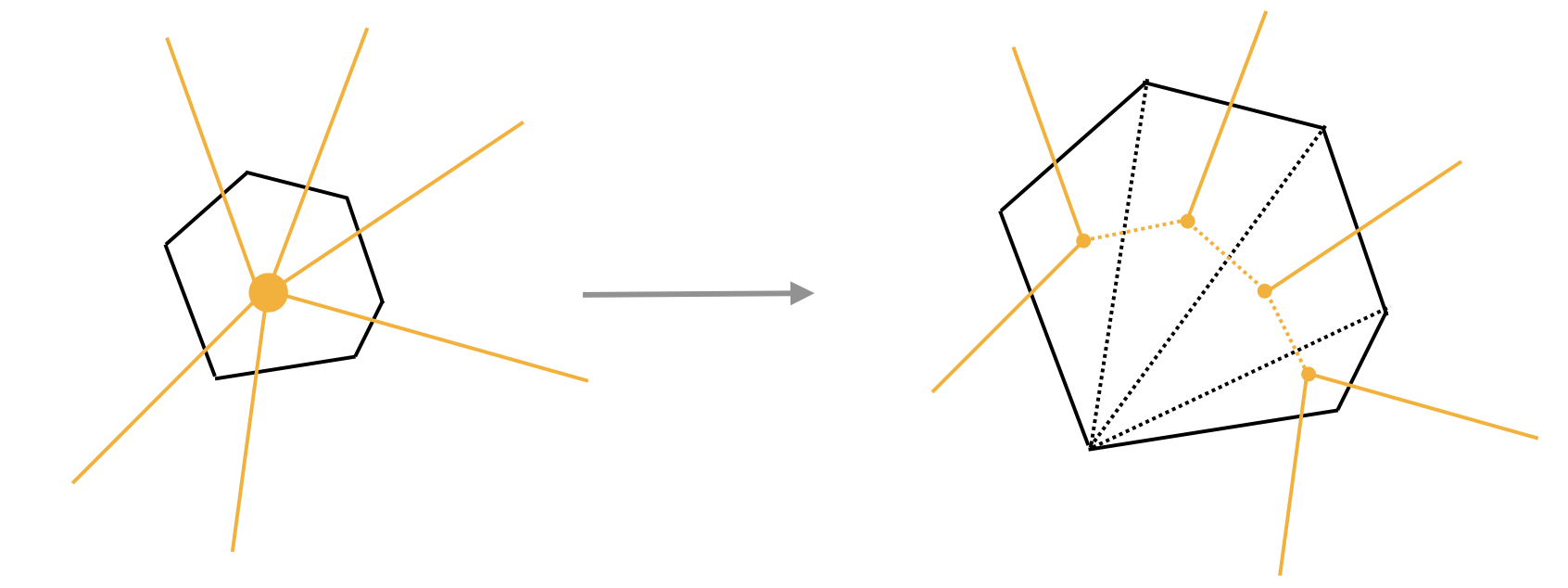}
\caption{
Unfolding a higher valent graph node into a 3-valent tree by triangulating its dual polygon. One can then apply the geometric formula for Ising zeros on the triangulation and generalize it to graphs of arbitrary valence, as shown in \cite{Bonzom:2024zka}.
\label{fig:unfolding1}}
\end{figure}

\begin{figure}[ht!]
\centering
\includegraphics[width=40mm]{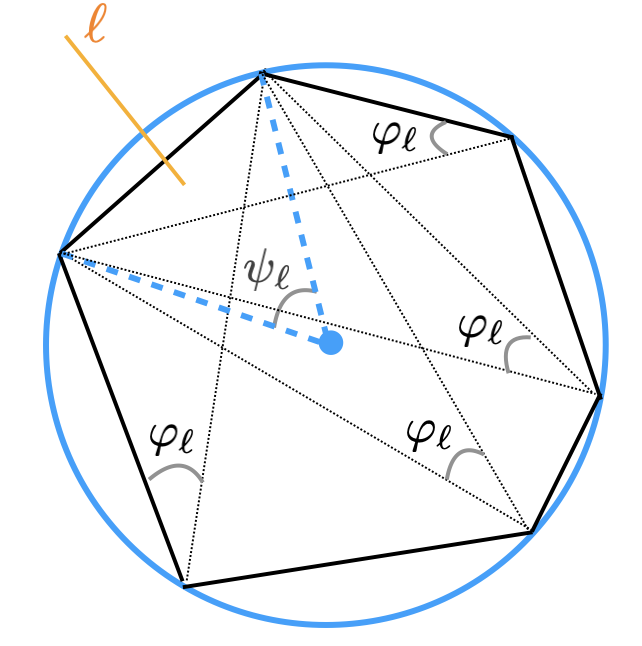}
\caption{
If a polygon is inscribed in a circle, the opposite triangle angle of a given edge $\varphi_{\ell}$ is always the same whatever the chosen opposite summit, and is automatically equal to half of the center angle $\psi_{\ell}$.
%
Thus working on embedded circle patterns, one gets  the generalized formula \eqref{eqn:circlepattern} for higher valent graphs  from the geometric ansatz \eqref{eq_Ising_zeroes} for 3-valent graphs and their dual 2d triangulations, as shown in \cite{Bonzom:2024zka}.
\label{fig:unfolding2}}
\end{figure}

This formula was further generalized in \cite{Bonzom:2024zka} to planar graphs with nodes of arbitrary valence in terms of the geometry of dual circle patterns, that is polygonal surfaces such that each polygonal face is inscribed in a circle. In this more general context, one still considers the opposite triangle angles for any triangulation of the polygonal faces. Indeed, the condition of being inscribed in a circle automatically ensures that those triangle angles are always equal, whatever the choice of triangulation, and actually equal to half the angle at the center of the circle, as depicted on fig.\ref{fig:unfolding1} and fig.\ref{fig:unfolding2}, yielding the formula:
\be
\label{eqn:circlepattern}
Y_{\ell}^{(0)}
=
e^{\pm i\f{\theta_{\ell}}2}\,
\sqrt{\tan\f{\vphi^{s}_{\ell}}2\,\tan\f{\vphi^{t}_{\ell}}2}
=
e^{\pm i\f{\theta_{\ell}}2}\,
\sqrt{\tan\f{\psi^{s}_{\ell}}4\,\tan\f{\psi^{t}_{\ell}}4}
\,,
\ee
where $\pm$ is an overall sign in front of all the dihedral angles.

This formula was proven explicitly for the tetrahedral graph in \cite{Bonzom:2019dpg} and was conjectured in the general case in \cite{Bonzom:2024zka}.
A general mathematically-rigorous proof of this geometric formula for Ising zeros in the general case is nevertheless lacking. Moreover, a few details of the formula still require work, e.g. a clear prescription for the sign of the dihedral angles and the square-root of the trigonometric functions, and the analytical continuation of the formulas to complex angles, which seems necessary to obtain all the Ising zeros, as underlined in \cite{Bonzom:2024zka}.

Here, we take special care in the definition of the dihedral angles, and provide a systematic choice of sign depending on the convexity or concavity of the triangulations. We then illustrate this choice by working out  analytically the case of a double pyramidal triangulations (dual to the cubic graph), and providing thorough numerical results supporting the formula by testing it on random  triangulations (with the topology of a 2-sphere).

\subsection{Dihedral angles, orientation and convexity}
\label{sec:orientation}


The original prescription from \cite{Bonzom:2019dpg} and \cite{Bonzom:2024zka}, as given above in \eqref{eq_Ising_zeroes} and \eqref{eqn:circlepattern}, was slightly imprecise in defining the sign of the dihedral angles $\theta_{\ell}$ entering the complex phase factors in the zeros' ansatz. Since the loop polynomial is holomorphic in the couplings $Y_{\ell}$'s, it is obvious that the complex conjugate of a root is again a root. So, simultaneously switching the signs of all the dihedral angles should still give a zero of the loop polynomial. This amounts to looking at the 2d triangulation from inside out. Indeed, since the 2d triangulation is planar, it surrounds a 3-ball in the ambient 3d space, and we can consider both inside or outside 3-balls. Nevertheless, we expect that switching the sign of individual dihedral angles should not yield other roots.
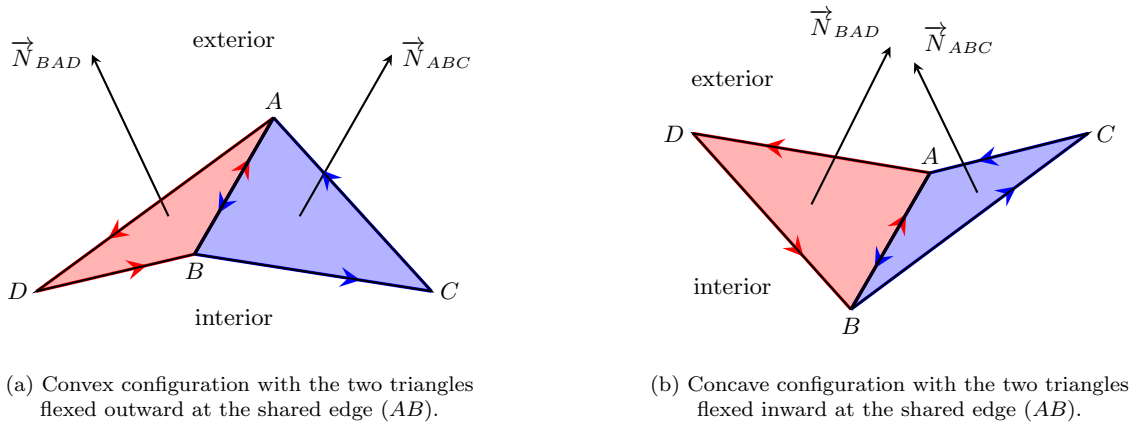
\begin{figure}
\centering
\subfloat[Convex configuration with the two triangles flexed outward at the shared edge $(AB)$.]{
\begin{tikzpicture}[scale=1.05]

\coordinate(a) at (0,2);
\coordinate(b) at (-1,.27);
\coordinate(c) at (2,-.2);
\coordinate(d) at (-3,-.2);
\coordinate(C) at (.33,.75);
\coordinate(D) at (-1.33,.75);
\coordinate(nC) at (1.5,2.8);
\coordinate(nD) at (-2.3,2.8);
\fill[fill=red,fill opacity=0.3]  (a)--(b)--(d)--(a);
\fill[fill=blue,fill opacity=0.3]  (a)--(c)--(b)--(a);

\draw (a) node[above]{$A$};
\draw (b) node[below]{$B$};
\draw (c) node[right]{$C$};
\draw (d) node[left]{$D$};

\draw[->, thick,>=stealth] (C)--(nC) node[right]{$\overrightarrow{N}_{ABC}$};
\draw[->, thick,>=stealth] (D)--(nD) node[left]{$\overrightarrow{N}_{BAD}$};


\draw[red, very thick, decoration={markings,mark=at position 0.7 with {\arrow[scale=1.5,>=stealth]{>}}},postaction={decorate}] (b)--(a);
\draw[red, very thick, decoration={markings,mark=at position 0.7 with {\arrow[scale=1.5,>=stealth]{>}}},postaction={decorate}] (a)--(d);
\draw[red, very thick, decoration={markings,mark=at position 0.7 with {\arrow[scale=1.5,>=stealth]{>}}},postaction={decorate}] (d)--(b);
\draw[blue, very thick, decoration={markings,mark=at position 0.7 with {\arrow[scale=1.5,>=stealth]{>}}},postaction={decorate}] (b)--(c);
\draw[blue, very thick, decoration={markings,mark=at position 0.7 with {\arrow[scale=1.5,>=stealth]{>}}},postaction={decorate}] (a)--(b);
\draw[blue, very thick, decoration={markings,mark=at position 0.7 with {\arrow[scale=1.5,>=stealth]{>}}},postaction={decorate}] (c)--(a);
\draw[thick] (b)--(c)--(a);
\draw[thick] (b)--(d)--(a);
\draw[very thick] (a)--(b);


\draw (-0.5,3) node{exterior};
\draw (b)++(0.5,-.8) node{interior};
\draw (0.5,-1) node{};

\end{tikzpicture}
}
\hspace{20mm}
\subfloat[Concave configuration with the two triangles flexed inward at the shared edge $(AB)$.]{
\begin{tikzpicture}[scale=1.05]

\coordinate(a) at (1,2);
\coordinate(b) at (0,.27);
\coordinate(c) at (3,2.5);
\coordinate(d) at (-2,2.5);
\coordinate(C) at (1.6,1.75);
\coordinate(D) at (-.5,1.6);
\coordinate(nC) at (.8,3.4);
\coordinate(nD) at (.5,3.6);
\fill[fill=red,fill opacity=0.3]  (a)--(b)--(d)--(a);
\fill[fill=blue,fill opacity=0.3]  (a)--(c)--(b)--(a);

\draw (a) node[above]{$A$};
\draw (b) node[below]{$B$};
\draw (c) node[right]{$C$};
\draw (d) node[left]{$D$};

\draw[red, very thick, decoration={markings,mark=at position 0.7 with {\arrow[scale=1.5,>=stealth]{>}}},postaction={decorate}] (b)--(a);
\draw[red, very thick, decoration={markings,mark=at position 0.7 with {\arrow[scale=1.5,>=stealth]{>}}},postaction={decorate}] (a)--(d);
\draw[red, very thick, decoration={markings,mark=at position 0.7 with {\arrow[scale=1.5,>=stealth]{>}}},postaction={decorate}] (d)--(b);
\draw[blue, very thick, decoration={markings,mark=at position 0.7 with {\arrow[scale=1.5,>=stealth]{>}}},postaction={decorate}] (b)--(c);
\draw[blue, very thick, decoration={markings,mark=at position 0.7 with {\arrow[scale=1.5,>=stealth]{>}}},postaction={decorate}] (a)--(b);
\draw[blue, very thick, decoration={markings,mark=at position 0.7 with {\arrow[scale=1.5,>=stealth]{>}}},postaction={decorate}] (c)--(a);
\draw[thick] (b)--(c)--(a);
\draw[thick] (b)--(d)--(a);
\draw[very thick] (a)--(b);

\draw (-1.5,3.2) node{exterior};
\draw (b)++(-1.5,0.3) node{interior};
\draw (-1.5,-0.3) node{};

\draw[->, thick,>=stealth] (C)--(nC) node[above right]{$\overrightarrow{N}_{ABC}$};
\draw[->, thick,>=stealth] (D)--(nD) node[above left]{$\overrightarrow{N}_{BAD}$};

\end{tikzpicture}
}
\caption{
With an anti-clockwise orientation around each triangle, we define the outward normal vectors $\protect\overrightarrow{N}_{ABC}$ and $\protect\overrightarrow{N}_{BAD}$.
The scalar product $\protect\overrightarrow{N}_{ABC}\cdot\protect\overrightarrow{N}_{BAD}$ gives the cosine $\cos\theta_{AB}$ of the dihedral angle.
Considering the shared edge $\protect\overrightarrow{AB}$, we choose as first triangle the one that has the same orientation as the chosen shared vector - here the triangle (ABC). With this convention, we then consider the cross product $(\protect\overrightarrow{N}_{ABC}\w \protect\overrightarrow{N}_{BAD})$. It is automatically collinear to  $\protect\overrightarrow{AB}$. If it is oriented in the same direction as $\protect\overrightarrow{AB}$, we have identified a convex configuration (on the left). Otherwise, it is a concave configuration (on the right).
%
\label{fig:normal}
\label{fig:normals}
}
\end{figure}

Indeed, we consider a 2d triangulation, embedded  in the flat 3d space, with the topology of a 2-sphere. It is the boundary of a 3-ball, and we can clearly distinguish the interior and exterior region. So for every triangle, we define the outward-oriented normal vector to the triangle. More precisely, since the surface has the topology of a 2-sphere, we can choose an anti-clock-wise orientation around every triangle of the 2d triangulation, that is a cyclic ordering of the vertices around each triangle (such that it is anti-clock-wise if we look at the 2d triangulation from the exterior). Let us then consider a triangle with ordered vertices $A,B,C$. Its outward normal vector is, as depicted on fig.\ref{fig:normal},
\be
\overrightarrow{N}_{ABC}
=\f12\overrightarrow{AB}\w\overrightarrow{AC}
=\f12\overrightarrow{BC}\w\overrightarrow{BA}
=\f12\overrightarrow{CA}\w\overrightarrow{CB}
\,.
\ee
Now let us consider the neighboring triangle sharing the edge $AB$ and call its third vertex $D$. The anti-clockwise order around that triangle is $B,A,D$, so its outward normal vector is:
\be
\overrightarrow{N}_{BAD}
=\f12\overrightarrow{BA}\w\overrightarrow{BD}
=\f12\overrightarrow{AD}\w\overrightarrow{AB}
=\f12\overrightarrow{DB}\w\overrightarrow{DA}
\,.
\ee
The scalar product between the two normal vectors gives the cosine of the dihedral angle between the two triangles:
\be
\overrightarrow{N}_{ABC}\cdot\overrightarrow{N}_{BAD}
=
|N_{ABC}|\,
|N_{BAD}|\,
\cos\theta_{AB}
\,,
\ee
where the norms $|N_{ABC}|$ and $|N_{BAD}|$ give the respective areas of the triangles, and $\theta_{AB}$ is the dihedral angle around the edge $(AB)$. This scalar product does not allow to decide of the sign of $\theta_{AB}$. To choose that sign, we need access to $\sin\theta_{AB}$, which would be given by the cross product of the two normal vectors. Thus our sign prescription for the dihedral angle is to always take it positive, $\theta_{AB}\in[0,\pi]$, and to adjust the sign of the complex phase factor in the coupling $Y_{AB}$ according to the direction of $\overrightarrow{N}_{ABC}\w \overrightarrow{N}_{BAD}$, as illustrated on fig.\ref{fig:normals}:
\begin{itemize}
\item if $(\overrightarrow{N}_{ABC}\w \overrightarrow{N}_{BAD})\cdot\overrightarrow{AB}\,>0$, the two triangles are flexed outward, this is a convex configuration and we take the phase factor $\exp(+ i \theta_{AB}/2)$;

\item if $(\overrightarrow{N}_{ABC}\w \overrightarrow{N}_{BAD})\cdot\overrightarrow{AB}\,<0$, the two triangles are flexed inward, this is a concave configuration and we take the phase factor $\exp(- i \theta_{AB}/2)$.

\end{itemize}
We check this sign prescription analytically in the example of the double pyramid below in the section \ref{sec:doubleP}, and numerically for random triangulations in section \ref{sec:numerics}. These validates our sign prescription and also show that this provides an indirect test of the convexity of the surface. 

\subsection{Examples: Triangle and Tetrahedron}

\begin{figure}[ht!]
\centering
\begin{tikzpicture}[scale=.8]

\coordinate (A) at (-1,0);
\coordinate (B) at (1.5,0);
\centerarc[](0,0)(90:270:1);
\centerarc[](.5,0)(-90:90:1);

\draw (A) node{$\bullet$}; 
\draw (B) node{$\bullet$}; 
\draw (A)--node[above]{$2$} (B);
\draw (0,1)--node[above]{$1$} (.5,1);
\draw (0,-1)--node[above]{$3$} (.5,-1);
\end{tikzpicture}
\hspace*{30mm}
\begin{tikzpicture}[scale=.7]

\coordinate (a) at (4,-1.4);
\coordinate (b) at (6.3,-1.4);
\coordinate (c) at (4.7,1.4);

\draw (a)--node[below]{$1$}(b)--node[right]{$2$}(c)--node[left]{$3$}(a);

\centerarc[](a)(0:74:.4);
\centerarc[](b)(120:180:.4);
\centerarc[](c)(254:300:.4);
\draw (c)++(0.15,-.75) node{$\vphi_{1}$}; 
\draw (a)++(.7,.3) node{$\vphi_{2}$}; 
\draw (b)++(-.6,.35) node{$\vphi_{3}$}; 
\end{tikzpicture}
\hspace*{30mm}
\begin{tikzpicture}[scale=.7]

\coordinate (x) at (8,0);
\coordinate (y) at (9,.8);
\coordinate (z) at (11.5,.2);
\draw (x)--(y)--(z)--(x);
\fill[fill=black,fill opacity=0.05]  (x)--(y)--(z)--(x);
\coordinate (C) at (9.5,.4);
\draw (C) node{$\bullet$}; 
\coordinate (D) at (9,2);
\coordinate (E) at (9.97,.-1.2);

\draw[->, thick,>=stealth](C)--(D) ;
\draw[->, thick,>=stealth](C)--(E) ;

\end{tikzpicture}
\caption{
On the left, the $\Theta$ graph with two trivalent nodes and three links between them.
On the right, the dual geometry of the $\Theta$ graph, as a triangular pancake, with two identical triangles glued on top of each other but pointing in opposite direction, so that the dihedral angle around each edge is a full 180 degrees.
In the middle, the geometry of the triangle with the three angles $\vphi_{k}$ opposite to the three edges.
\label{fig:Thetagraph}}
\end{figure}
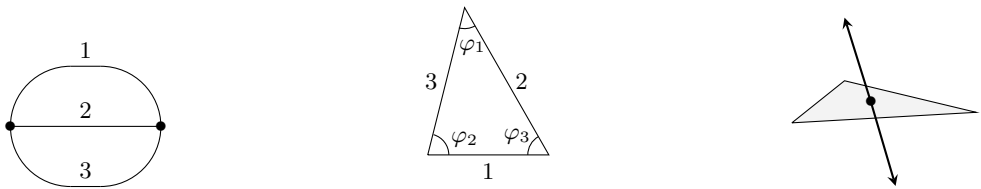

The first example, as introduced in \cite{Bonzom:2019dpg}, is the simplest trivalent graph -the $\Theta$ graph- made of 2 vertices with three links between them, as depicted on fig.\ref{fig:Thetagraph}.
Numbering the three links $1,2,3$, the $\Theta$ graph contains three non-trivial cycles, and its loop polynomial read:
\be
P_{\Theta}[Y_{1},Y_{2},Y_{3}]
=
1+Y_{1}Y_{2}+Y_{2}Y_{3}+Y_{3}Y_{1}
\,.
\ee

Since each vertex is dual to a triangle, the dual 2d triangulation of the $\Theta$ graph is a triangular pancake, with two identical triangles, glued together with opposite normals, so that the dihedral angles around each edge is $\pi$. The corresponding geometric ansatz is thus:
\be
Y_{1}^{(0)}
=
i\tan\f{\vphi_{1}}2
\,\quad
Y_{2}^{(0)}
=
i\tan\f{\vphi_{2}}2
\,\quad
Y_{3}^{(0)}
=
i\tan\f{\vphi_{3}}2
\,
\ee
where the $\vphi_{k}$ are the opposite triangle angles and $i=\exp(i\pi/2)$ gives the complex phase factors encoding the dihedral angles, as illustrated on fig.\ref{fig:Thetagraph}. Then $P_{\Theta}[Y_{1}^{(0)},Y_{2}^{(0)},Y_{3}^{(0)}]$ vanishes as soon as the sum of the three angles is equal to $\pi$:
\be
\vphi_{1}+\vphi_{2}+\vphi_{3}=\pi
\quad\Rightarrow\quad
\tan\f{\vphi_{1}}2\tan\f{\vphi_{2}}2+\tan\f{\vphi_{2}}2\tan\f{\vphi_{3}}2+\tan\f{\vphi_{3}}2\tan\f{\vphi_{1}}2=1
\quad\Leftrightarrow\quad
P_{\Theta}[Y_{1}^{(0)},Y_{2}^{(0)},Y_{3}^{(0)}]=0
\,.
\ee
We recognize the law of (half-)tangents for the angles of a triangle.

\medskip

\begin{figure}[ht!]
\centering
\subfloat[Tetrahedral graph $\cT=T^{*}$.]{
\begin{tikzpicture}[scale=1.43]

\coordinate(DD) at (4.5,1.5);
\coordinate(CC) at (6,.8);
\coordinate(BB) at (5,2.5);
\coordinate(AA) at (6.5,1.5);
\draw (AA) node {$\bullet$} ++(.2,0) node{$a$} ;
\draw (BB) node {$\bullet$}++(0,.22) node{$b$};
\draw (CC) node {$\bullet$}++(.05,-.18) node{$c$} ;
\draw (DD) node {$\bullet$}++(-.25,0) node{$d$};
\draw (AA)--node[above right]{$Y_{4}$}(BB);
\draw (AA)--node[below right]{$Y_{5}$}(CC);
\draw (AA)--(DD);
\draw  (5.55,1.9) node {$Y_{6}$};
\draw (BB)--(CC);
\draw  (5,1.67) node {$Y_{3}$};
\draw (BB)--node[above left]{$Y_{2}$}(DD);
\draw (CC)--node[below left]{$Y_{1}$}(DD);

\end{tikzpicture}
}
\hspace{30mm}
\subfloat[Dual tetrahedron $T=\cT^{*}$\!.]{
\begin{tikzpicture}[scale=1.12]

\coordinate(a) at (0,0) ;
\coordinate(ab) at (1.1,-0.35) ;
\coordinate(b) at (2.2,-.7);
\coordinate(bc) at (1.6,1.15) ;
\coordinate(ac) at (.5,1.5);
\coordinate(c) at (1,3);
\coordinate(ad) at (1.75,.15);
\coordinate(bd) at (2.85,-.2);
\coordinate(cd) at (2.25,1.65);
\coordinate(d) at (3.5,0.3);

\coordinate(C) at (2.15,-.2);
\draw[orange, thick] (C)--(ab);
\draw[orange, thick] (C)--(bd);
\draw[orange, thick] (C)--(ad);

\coordinate(A) at (2.1,1);
\draw[orange, thick] (A)--(bc);
\draw[orange, thick] (A)--(cd);
\draw[orange, thick] (A)--(bd);

\coordinate(B) at (1.2,1.5);
\draw[orange, very thick,dotted] (B)--(ac);
\draw[orange, very thick,dotted] (B)--(ad);
\draw[orange, very thick,dotted] (B)--(cd);

\coordinate(D) at (1,1);
\draw[orange, thick] (D)--(ab);
\draw[orange, thick] (D)--(bc);
\draw[orange, thick] (D)--(ac);

\draw (a)--(b)--(c)--(a);
\draw[dotted] (a)--(d);
\draw (b)--(d)--(c);

\draw (A) node[orange] {$\bullet$} ++(.25,-0.05) node{$a$} ;
\draw (B) node[orange] {$\bullet$}++(0,.24) node{$b$};
\draw (C) node[orange] {$\bullet$}++(.1,-.18) node{$c$} ;
\draw (D) node[orange] {$\bullet$}++(-.25,-.1) node{$d$};

\draw(ab)node[below]{$Y_{1}$} ;
\draw(ac)node[left]{$Y_{2}$} ;
\draw(bc)++(0.25,.1)node{$Y_{3}$} ;
\draw(ad)++(-0.3,.2) node{$Y_{6}$} ;
\draw(bd)node[right]{$Y_{5}$} ;
\draw(cd)node[above]{$Y_{4}$} ;

\end{tikzpicture}
}
\caption{
From tetrahedral graph to its dual tetrahedron: each graph node $(a,b,c,d)$ becomes a triangle, and each link $(1,..,6)$ becomes an edge shared by two triangles. The geometric coupling $Y_{\ell}$ involves the dihedral angle between the two triangles sharing the edge dual to the link $\ell$, and as well as the angles within those triangles.
\label{fig:tetrahedron}}
\end{figure}
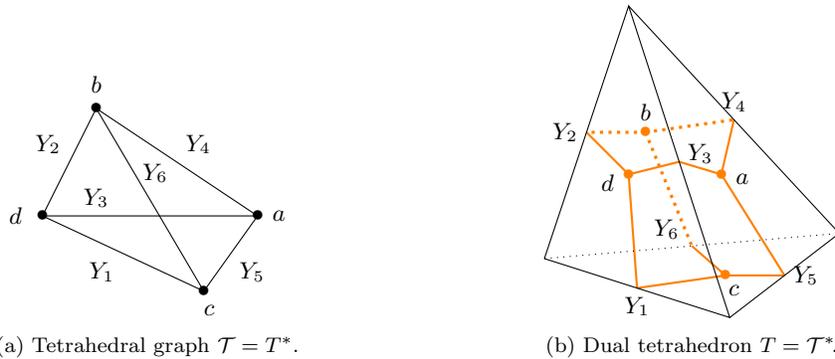

The next case for a trivalent graph is the tetrahedral graph, whose dual is a tetrahedron, as illustrated on fig.\ref{fig:tetrahedron}. We call the tetrahedron $T$ and the tetrahedral graph $\cT=T^{*}$. We have four three-cycles (going around the four summits of the tetrahedron) and three four-cycles (corresponding to the parallelograms linking the mid points of pairs of opposite edges), giving the following loop polynomial on the tetrahedral graph :
\be
P_{\cT}[Y_{1},..,Y_{6}]
=
1+Y_{1}Y_{2}Y_{6}+Y_{1}Y_{5}Y_{3}+Y_{4}Y_{2}Y_{3}+Y_{4}Y_{5}Y_{6}
+Y_{1}Y_{2}Y_{4}Y_{5}+Y_{2}Y_{3}Y_{5}Y_{6}+Y_{1}Y_{3}Y_{4}Y_{6}
\,.
\ee

The general case for the tetrahedron is worked out in full details in \cite{Bonzom:2019dpg}. Here we simply review the special case with homogeneous couplings $Y_{1}=..=Y_{6}=Y$, for which the loop polynomial reduces to:
\be
P_{\cT}^{homo}[Y]
=
1+4Y^{3}+3Y^{4}
=
3(Y+1)^{2}\left(Y-\f13(1-i\sqrt{2})\right)\left(Y-\f13(1+i\sqrt{2})\right)
\,.
\ee
The $-1$ root corresponds to a totally squashed degenerate tetrahedron, while the other roots are the geometric couplings for the equilateral tetrahedron:
\be
Y^{(0)}_{\cT}
=
e^{i\f{\theta_{0}}2}\tan\f\pi6
=
\f13(1+i\sqrt{2})
\,,\qquad
\overline{Y^{(0)}_{\cT}}
=
e^{-i\f{\theta_{0}}2}\tan\f\pi6
=
\f13(1-i\sqrt{2})
\,,
\ee
where $\theta_{0}$ is the dihedral angle between triangles in the equilateral tetrahedron and satisfies $\cos\theta_{0}=-\f13$.

An interesting point is that this perfectly illustrates the high temperature - low temperature duality of the 2d Ising model. As explained in details in \cite{baxter}, both the high temperature and low temperature expansions of the 2d Ising partition function can be written in terms of sum over loops, but on dual graphs\footnotemark{}. As pointed out in \cite{Bonzom:2019dpg}, the tetrahedral graph is dual to a tetrahedron, so it should be self-dual under the map between couplings at low T and hight T. More precisely, the map reads:
\be
\cD[Y]=\f{1-Y}{1+Y}\,.
\ee
\footnotetext{
A first expansion of the Ising partition function is the sum over even subgraphs, that we used here, obtained by linearizing the exponential Ising weights:
\be
Z_{\Gamma}[\{y_{\ell}\}]
=
\sum_{\{\sigma_{n}=\pm\}}\prod_{\ell} e^{y_{\ell}\sigma_{s(\ell)}\sigma_{t(\ell)}}
=
\left(\prod_{\ell}\cosh y_{\ell}\right)\,
\sum_{\{\sigma_{n}=\pm\}} \prod_{\ell}\,(1+\sigma_{s(\ell)}\sigma_{t(\ell)}\tanh y_{\ell})
=
2^{N_{n}}\prod_{\ell}
\cosh y_{\ell}\,
\sum_{\cG_{\textrm{even}}\subset \Gamma} \,\,\prod_{\ell\in\cG}\tanh y_{\ell}
\,,
\nn
\ee
where we sum over all even subgraphs, including the empty set and the original graph $\Gamma$ itself. The even subgraph condition comes from combining the factors $\sigma_{s(\ell)}\sigma_{t(\ell)}$ in such a way that each $\sigma_{n}$ appears an even number of times for every node $n$, so that the sum over $\sigma_{n}=\pm$ does not vanish. Re-introducing the temperature in the couplings, $y_{\ell}=\beta j_{\ell}$ with the inverse temperature $\beta$ and bare Ising couplings $j_{\ell}$, one sees that the effective couplings of this expansion $Y_{\ell}=\tanh y_{\ell}$ are very small for high temperature as $\beta\rightarrow 0$, making it a perturbative expansion at high T. \\
A second expansion is given by the cluster decomposition for a planar graph: we draw the boundary between positive spins $\sigma_{n}=+1$ and negative spins $\sigma_{n}=-1$. This boundary is an even subgraph of the dual graph $\Gamma^{*}$, whose nodes are dual to $\Gamma$'s faces and whose faces are dual to $\Gamma$'s nodes. Dual links are in one-to-one correspondance to the original links. Since the Ising weight between two identical spins gives $e^{+y_{\ell}}$, while it gives $e^{-y_{\ell}}$ when the two spins are different, we can set all the weights to $e^{+y_{\ell}}$ by default and adding a penalty factor $e^{-2y_{\ell}}$ for all links on the boundary between clusters of positive spins and negative spins. This yields:
\be
Z_{\Gamma}[\{y_{\ell}\}]
=
\prod_{\ell}
e^{+y_{\ell}}\,
\sum_{\cG^{*}_{\textrm{even}}\subset \Gamma^{*}} \,\,\prod_{\ell\in\cG^{*}}e^{-2y_{\ell}}
\,,
\nn
\ee
The effective coupling $Y^{*}_{\ell} =e^{-2y_{\ell}}=e^{-2\beta j_{\ell}}$ is now very small for low T as $\beta\rightarrow +\infty$, making this formula a low temperature perturbative expansion.
The map between the two expansions is given by the definition of the hyperbolic tangent, $Y_{\ell}=(1+Y^{*}_{\ell})/(1-Y^{*}_{\ell})$.
}
This is clearly an involution and maps the loop polynomial $P_{T^{*}}$ of the tetrahedral graph onto the loop polynomial of its dual tetrahedron:
\beq
P_{\cT}\Big{[}\cD[Y_{1}],..,\cD[Y_{6}]\Big{]}
&=&
\f8{\prod_{\ell=1}^{6}(1+Y_{\ell})}
\big{(}
1+Y_{1}Y_{2}Y_{3}+Y_{1}Y_{5}Y_{6}+Y_{4}Y_{5}Y_{3}+Y_{4}Y_{2}Y_{6}
+Y_{1}Y_{2}Y_{4}Y_{5}+Y_{2}Y_{3}Y_{5}Y_{6}+Y_{1}Y_{3}Y_{4}Y_{6}
\big{)}
\nn
\\
&=&
\f8{\prod_{\ell=1}^{6}(1+Y_{\ell})}
P_{\cT}[Y_{4},Y_{5},Y_{6},Y_{1},Y_{2},Y_{3}]
=
\f8{\prod_{\ell=1}^{6}(1+Y_{\ell})}
P_{T}[Y_{1},Y_{2},Y_{3},Y_{4},Y_{5},Y_{6}]
\,.
\eeq
This implies a self-duality relation in the homogeneous coupling case:
\be
P_{\cT}^{homo}\Big{[}\cD[Y]\Big{]}
=
\f8{(1+Y)^{6}}
P_{\cT}^{homo}[Y]
\,,
\ee
which means that the dual of a root, as long as it is different from $-1$, under $\cD$ is again a root of the loop polynomial. Here, we indeed have:
\be
\cD[Y^{(0)}_{\cT}]=\overline{Y^{(0)}_{\cT}}
\,,\qquad
\cD[\overline{Y^{(0)}_{\cT}}]=Y^{(0)}_{\cT}
\,.
\ee
As originally introduced by Kramers and Wannier \cite{PhysRev.60.252}, and reviewed in \cite{baxter}, this high T/low T duality may be used to identify critical couplings (for instance, for the square lattice). Here we see the complex phase factors, due to the dihedral angles of the 2d triangulation,  are crucial to ensure the duality applies to the geometric formula for the zeros of the 2d Ising model on finite graphs.

\subsection{Example:  the Pyramid}

Let us consider a simple configuration beyond triangulations: the pyramid with square base. It is also especially interesting because the pyramidal graph is self-dual, just like the tetrahedral graph, in the sense that its dual graph is again a pyramidal graph. This will allow to check the high T/low T duality.
\begin{figure}
\centering
\subfloat[Square base pyramid of height $h$ and with triangle angles $\vphi$ and $\tvphi$.]{
\begin{tikzpicture}[scale=1.7]

\coordinate(A) at (0,0);
\coordinate(B) at (2,0);
\coordinate(C) at (.7,.7);
\coordinate(D) at (2.7,.7);
\coordinate(S) at (1.3,2.5);
\coordinate(s) at (1.3,.35);

\draw (A)--node[below]{$Y$} (B)--node[right]{$Y$}(D);
\draw[thick, dotted] (D)--(C)--node[right]{$Y$}(A);
\draw (A)--node[left]{$\tY$}(S);
\draw (B)--node[right]{$\tY$}(S);
\draw[thick, dotted] (C)--(S);
\draw (D)--node[right]{$\tY$}(S);
\draw[blue] (s)--node[right]{$h$} (S);
\draw[blue] (1.3,.45)--(1.4,.45)--(1.4,.35);

\draw(C)++(.4,.7) node{$\tY$} ;
\draw(C)++(.85,.15) node{$Y$} ;

\centerarc[](B)(45:106:.2);
\centerarc[](S)(-75:-52:.4);
\draw(B)++(.1,.32) node{$\tvphi$} ;
\draw(S)++(.22,-.45) node{$\vphi$} ;

\end{tikzpicture}
}
\hspace{25mm}
\subfloat[Pyramidal graph: faces of the pyramid become graph nodes, linked when the faces are adjacent.]{
\begin{tikzpicture}[scale=1.7]

\coordinate(A) at (0,0);
\coordinate(B) at (2,0);
\coordinate(C) at (.7,.5);
\coordinate(D) at (2.7,.5);
\coordinate(S) at (1.3,-1.8);
\draw (A) node {$\bullet$};
\draw (B) node {$\bullet$};
\draw (C) node {$\bullet$};
\draw (D) node {$\bullet$};
\draw (S) node {$\bullet$};

\draw (A)--node[above]{$\tY$} (B)--node[left]{$\tY$}(D)--node[above]{$\tY$}(C)--node[left]{$\tY$}(A);
\draw (A)--node[left]{$Y$}(S);
\draw (B)--node[left]{$Y$}(S);
\draw[thick, dotted] (C)--node[right]{$Y$}(S);
\draw (D)--node[right]{$Y$}(S);
\end{tikzpicture}
}
\caption{
The square base pyramid and the pyramidal graph are dual to each other: they have the same cellular structure but the role of summit  and base is inverted.
\label{fig:pyramid}}
\end{figure}
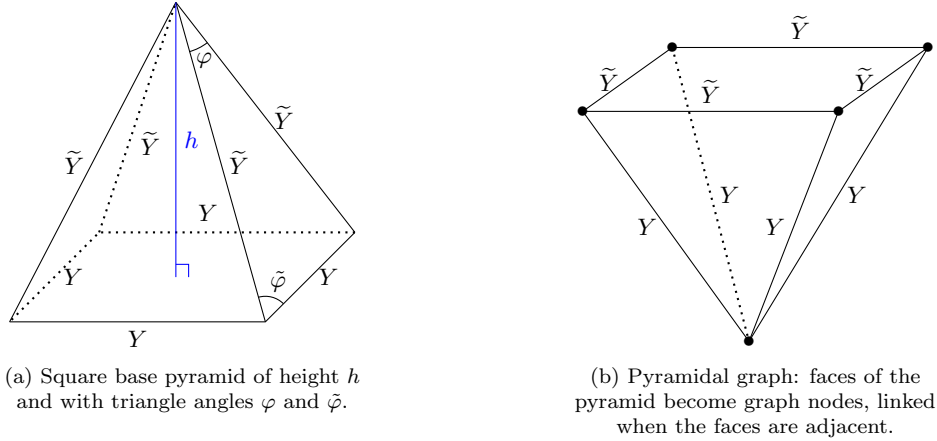

So, in more details, we consider a square base with edge length 2 and a summit of height $h$ centered above the square, as illustrated on fig.\ref{fig:pyramid}. So the four faces (on top of the square) are isosceles triangles, with edge length $\sqrt{h^{2}+2}$. This allows to compute its summit angle $\vphi$ and base angles $\tvphi$. More precisely, one can directly compute their half-tangent, accordingly to the well-known trigonometric formula given in fig.\ref{fig:halfangletan},
\be
\tan \f\vphi2=\f1{\sqrt{h^{2}+1}}
\,,\qquad
\tan \f\tvphi2=\f{\sqrt{h^{2}+2}-1}{\sqrt{h^{2}+1}}
\,.
\ee
Computing the scalar product between the outward normals to the faces, we get the cosine of the dihedral angles, $\theta$ around the base edges and $\ttheta$ around the summit edges,
\be
\cos\theta=\f{-1}{\sqrt{h^{2}+1}}
\,,\qquad
\cos\ttheta=\f{1}{h^{2}+1}
\,,
\ee
from which we deduce the corresponding complex phases:
\be
e^{i\theta}=\f{-1+ih}{\sqrt{h^{2}+1}}
\,,\qquad
e^{i\f\ttheta2}
=
\sqrt{\f{1+\cos\ttheta}2}+i\sqrt{\f{1-\cos\ttheta}2}
=
\f{\sqrt{h^{2}+2}+ih}{\sqrt{2(h^{2}+1)}}
\,.
\ee
\begin{figure}[h!]

\centering
\begin{tikzpicture}[scale=1.5]

\coordinate(x) at (0,0) ;
\coordinate(y) at (2,0);
\coordinate(z) at (1.3,1);
\draw (1.3,1.2) node{};

\draw[thick] (x)-- node[below]{$l_{1}$}(y)-- node[above right]{$l_{3}$}(z)--node[above left]{$l_{2}$}(x);
\centerarc[](x)(0:38:.4);
\draw (0.6,0.15) node{$\vphi_{12}$};
\draw (5.5,.5) node{\large $\tan^{2}\f{\vphi_{12}}2=\f{l_{3}^{2}-(l_{1}-l_{2})^{2}}{(l_{1}+l_{2})^{2}-l_{3}^{2}}
=
\f{(l_{3}+l_{1}-l_{2})(l_{3}+l_{2}-l_{1}))}{(l_{1}+l_{2}+l_{3})(l_{1}+l_{2}-l_{3})}$.};
\end{tikzpicture}

\caption{
Half-angle tangent formula.
\label{fig:halfangletan}}
\end{figure}

Since the base square can obviously be inscribed in a circle, we can apply the geometric ansatz for Ising zeroes:
\be
\label{eq:pyraYgeom}
\textrm{for base edges:}\quad
\big{(}Y^{(0)}\big{)}^{2}=e^{i\theta}\tan\f\pi8\tan\f\vphi2
=
\f{(-1+ih)(\sqrt{2}-1)}{h^{2}+1}
\,,
\ee
\be
\textrm{for summit edges:}\quad
\tY^{(0)}=e^{i\f\ttheta2}\tan\f\tvphi2
=
\f{(\sqrt{h^{2}+2}+ih)(\sqrt{h^{2}+2}-1)}{(h^{2}+1)\sqrt{2}}
\,,
\ee
where we have only computed the square of the coupling for base edges, because only this combination enters the loop polynomial.
Indeed, a simple counting of the cycles of the pyramidal graph gives its loop polynomial in terms of the two couplings $Y$ (between base and triangle) and $\tY$ (between triangles), which can be written in a compact way as:
\be
P_{pyra}[Y,\tY]
=
1+\tY^{4}+4\tY Y^{2}(1+\tY^{2})+2\tY^{2} Y^{2}(2+Y^{2})
\,.
\ee
In order to evaluate the value of this polynomial on the geometric ansatz $P_{pyra}[Y^{(0)},\tY^{(0)}]$, we notice that we can set all the terms of this polynomial on the same common denominator $(h^2+1)^4\sqrt{2}$. Then the numerator is a polynomial in $h$ with some square-root factors $\sqrt{2}$ and $\sqrt{h^{2}+2}$. We can thus re-group it in four packets, with (quartic) polynomial factors in $h$  in front of the terms in $1$, $\sqrt{2}$, $\sqrt{h^{2}+2}$ and $\sqrt{2}\sqrt{h^{2}+2}$.
A straightforward calculation\footnotemark{} of those four terms allows to check by hand that the geometric ansatz actually gives zeros of this polynomial whatever the value of the height $h$:
\be
\forall h\ge 0\,,\quad P_{pyra}[Y^{(0)},\tY^{(0)}]
=0
\,.
\ee
\footnotetext{
A simple consistency check is to look at the case $h\rightarrow \pm\infty$. In those asymptotic limit, $Y^{(0)}$ clearly vanishes while $\tY^{(0)}\rightarrow e^{\pm i\pi/4}$. Thus setting to zero the monomials with $Y^{(0)}$ factors, the loop polynomial reduces to $1+\tY^4$, which clearly vanishes for the asymptotic limit of $\tY^{(0)}$.
One can also consider the special value $h=0$, for which $(Y^{(0)})^{2}=1-\sqrt{2}$ and $\tY^{(0)}=\sqrt{2}-1$, so that the loop polynomial evaluates to:
\be
P_{pyra}[Y^{(0)},\tY^{(0)}]\Big{|}_{h=0}
=
1+(\sqrt{2}-1)^{4}
-4(\sqrt{2}-1)^{2}\big{(}1+(\sqrt{2}-1)^{2}\big{)}
-2(\sqrt{2}-1)^{3}\big{(}3-\sqrt{2}\big{)}
=
0\,.
\nn
\ee
}
A special case is the equilateral pyramid, for which summit and base edges have the same length. It is given by the height $h=\sqrt{2}$, for which the angles are given by:
\be
\textrm{for the equilateral pyramid:}\qquad
\vphi=\tvphi=\f\pi3
\,,\quad
\cos\theta=-\f1{\sqrt{3}}
\,,\quad
\cos\ttheta=\f1{3}
\,.
\ee
The corresponding geometric couplings are nevertheless not equal, but can be computed explicitly applying the formula above:
\be
(Y^{(eq)})^{2}=
\f13(\sqrt{2}-1)(-1+i\sqrt{2})
\,,\qquad
\tY^{(eq)}=
\f13(\sqrt{2}+i)
\,.
\ee
Finally, let us turn to the high T/low T duality. Applying the duality map $\cD$ to the couplings $Y$ and $\tY$ simply inverts the role of base and summit edges:
\be
P_{pyra}[\cD[Y],\cD[\tY]]
=
\f1{(1+Y)^{4}(1+\tY)^{4}}\,P_{pyra}[\tY,Y].
\ee
One can indeed check numerically that the dual of the geometric couplings, with inverted roles, are still roots of the loop polynomial. For instance, both pairs $(Y^{(eq)},\tY^{(eq)})$ and $(\cD[\tY^{(eq)}],\cD[Y^{(eq)}])$ are zeroes of $P_{pyra}$.
At this stage, we hit a puzzle: the dual of geometric couplings is again an Ising zero, but it is not a geometric coupling anymore, as already hinted in \cite{Bonzom:2019dpg}. Indeed, let us compute the dual of the summit edge coupling $\tY^{(eq)}$ of the equilateral case, and then square it in order to compare with the geometric ansatz for the base edge $Y^{(0)}$:
\be
\tY^{(eq)}=\f13(\sqrt{2}+i)
\quad\Rightarrow\quad
\cD[\tY^{(eq)}]=\f{1-i}{2+\sqrt{2}}
\quad\Rightarrow\quad
\big{(}\cD[\tY^{(eq)}]\big{)}^{2}=i(2\sqrt{2}-3)
\,.
\ee
This purely imaginary value can never be reached for any real value of the pyramid height by $(Y^{(0)})^{2}$ as given by \eqref{eq:pyraYgeom}. However, a quick calculation reveals that it does correspond to a complex value of the height $h$:
\be
h^{(eq)}_{dual}
=
-(1+\sqrt{2})+i
\,,
\ee
which also fits the value of $\cD[Y^{(eq)}]$.
This confirms the need to complexify the geometric formula for Ising zeros, as already pointed out in \cite{Bonzom:2024zka}. However, it is unclear how to complexify the geometry of 2d triangulations and what should be the appropriate prescription for complexified triangle and dihedral angles. It would nevertheless be enlightening, in the future, to understand the geometry underlying the high T/low T duality of the 2d Ising model, and why it has to be interpreted in terms of complex metrics.

\subsection{Examples: Cube and Double-Pyramid}

\begin{figure}
\centering
\begin{tikzpicture}[scale=1.6]

\coordinate(A) at (0,0);
\coordinate(B) at (2,0);
\coordinate(C) at (.7,.5);
\coordinate(D) at (2.7,.5);
\coordinate(S) at (1.3,1.7);
\coordinate(s) at (1.3,-1.3);

\coordinate(AB) at (1,0);
\coordinate(AC) at (.35,.25);
\coordinate(CD) at (1.35,0.5);
\coordinate(BD) at (2.35,.25);
\coordinate(SA) at (.65,.85);
\coordinate(SB) at (1.65,.85);
\coordinate(SC) at (1,1.1);
\coordinate(SD) at (2,1.1);
\coordinate(sA) at (.65,-.65);
\coordinate(sB) at (1.65,-.65);
\coordinate(sC) at (1,-.4);
\coordinate(sD) at (2,-.4);

\draw[very thick] (A)-- (B)--(D);
\draw[ultra thick, dotted] (D)--(C)--(A);
\draw[very thick] (A)--(S);
\draw[very thick] (B)--(S);
\draw[ultra thick, dotted] (C)--(S);
\draw[very thick] (D)--(S);
\draw[very thick] (A)--(s);
\draw[very thick] (B)--(s);
\draw[ultra thick, dotted] (C)--(s);
\draw[very thick] (D)--(s);

\draw (A) node {$\bullet$};
\draw (B) node {$\bullet$};
\draw (C) node {$\bullet$};
\draw (D) node {$\bullet$};
\draw (S) node {$\bullet$};
\draw (s) node {$\bullet$};

\draw[blue] (1.1,.65) node {$\bullet$};
\draw[blue] (1.2,-.55) node {$\bullet$};
\draw[blue] (2,.85) node {$\bullet$};
\draw[blue] (1.93,.-.35) node {$\bullet$};
\draw[blue] (SA)--(1.1,.65)--(SB)--(2,.85)--(SD);
\draw[blue] (1.1,.65)--(AB)--(1.2,-.55);
\draw[blue] (2,.85)--(BD)--(1.93,.-.35);
\draw[blue] (sA)--(1.2,-.55)--(sB)--(1.93,.-.35)--(sD);

\draw[opacity=.5,blue] (1.51,.93) node {$\bullet$};
\draw[opacity=.5,blue] (.65,.71) node {$\bullet$};
\draw[opacity=.5,blue] (.7,-.35) node {$\bullet$};
\draw[opacity=.5,blue] (1.51,.-.15) node {$\bullet$};
\draw[opacity=.5,blue] (SD)--(1.51,.93)--(SC)--(.65,.71)--(SA);
\draw[opacity=.5,blue] (.65,.71)--(AC)--(.7,-.35);
\draw[opacity=.5,blue] (sA)--(.7,-.35) --(sC)--(1.51,.-.15)--(sD);
\draw[opacity=.5,blue] (1.51,.-.15)--(CD)--(1.51,.93);

\end{tikzpicture}
\hspace{30mm}
\begin{tikzpicture}[scale=1.6]

\coordinate(A) at (0,0);
\coordinate(B) at (2,0);
\coordinate(C) at (.7,.5);
\coordinate(D) at (2.7,.5);
\coordinate(a) at (0,2);
\coordinate(b) at (2,2);
\coordinate(c) at (.7,2.5);
\coordinate(d) at (2.7,2.5);
\draw[blue,very thick] (A)--(B)--(D)--(d)--(b)--(a)--(A);
\draw[blue,very thick] (B)--(b);
\draw[blue,ultra thick, dotted] (C)--(c);
\draw[blue,very thick] (a)--(c)--(d);
\draw[blue,ultra thick, dotted] (A)--(C)--(D);

\draw[blue,thick] (A) node {$\bullet$};
\draw[blue,thick] (B) node {$\bullet$};
\draw[blue,thick] (C) node {$\bullet$};
\draw[blue,thick] (D) node {$\bullet$};
\draw[blue,thick] (a) node {$\bullet$};
\draw[blue,thick] (b) node {$\bullet$};
\draw[blue,thick] (c) node {$\bullet$};
\draw[blue,thick] (d) node {$\bullet$};

\draw (1,1) node {$\bullet$};
\draw (2.35,1.25) node {$\bullet$};
\draw (1.35,2.25) node {$\bullet$};
\draw[opacity=.5] (0.35,1.25) node {$\bullet$};
\draw[opacity=.5] (1.35,0.25) node {$\bullet$};
\draw[opacity=.5] (1.7,1.5) node {$\bullet$};
\draw (0,1)--(2,1)--(2.7,1.5);
\draw (1,0)--(1,2)--(1.7,2.5);
\draw (2.35,0.25)--(2.35,2.25)--(0.35,2.25);
\draw[opacity=.5] (1,0)--(1.7,0.5)--(1.7,2.5);
\draw[opacity=.5] (2.35,0.25)--(0.35,0.25)--(0.35,2.25);
\draw[opacity=.5] (0,1)--(.7,1.5)--(2.7,1.5);

\end{tikzpicture}
\caption{
Double pyramid and cube are dual to each other: faces become graph nodes and edges become dual links, and vice versa. On the left, the double pyramid geometry in black, with its dual cubic graph drawn in {\blue blue}. On the right, the cube geometry in {\blue blue}, with its dual double pyramidal graph drawn in black.
\label{fig:cube}}
\end{figure}
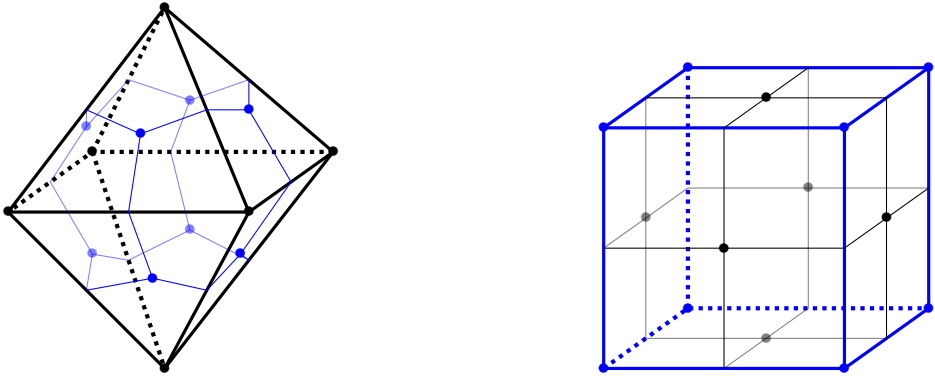

Another simple geometry to visualize is the regular cube, with its six squares and twelve edges. It is the dual of the double pyramid graph, as illustrated on fig.\ref{fig:cube}. Since each square face is inscribed in a circle, we can apply the geometric formula \eqref{eqn:circlepattern} for Ising zeros on circle patterns.

All the edges of the cube are equivalent, their dihedral angles are obviously $\pi/2$ and the opposite triangle angle are $\pi/4$, so that the geometric ansatz for the corresponding Ising zero gives:
\be
\cY^{(0)}=e^{i\f\pi4}\tan\f\pi8
=
\f{(1+i)}{\sqrt{2}}\,(\sqrt{2}-1)
\,.
\ee
Let us point out that the phase-less version $\tan\f\pi8=(\sqrt{2}-1)$  is the geometric coupling for the square lattice (in the thermodynamic limit of an infinite graph) and actually reproduces exactly the critical coupling for the square lattice.

Keep the convention that we refer to the loop polynomial by the graph on which the Ising model is defined, and not by the dual 2d geometry, the loop polynomial for the double-pyramidal graph for homogeneous coupling along the twelve links is:
\be
P_{doubleP}[\cY]
=
(1+\cY)^{4}(1+\cY^{2})^{2}\,(1-4\cY+8\cY^{2}-4\cY^{3}+\cY^{4})
\,.
\ee
Beside the trivial roots $-1$ and $\pm i$, the roots of this polynomial are $\cY^{(0)}$ and $\overline{\cY^{(0)}}$, confirming the geometric formula, plus another pair of roots $\cY^{(1)}$ and $\overline{\cY^{(1)}}$ with
\be
\cY^{(1)}=e^{i\f\pi4}\tan\f{3\pi}8
=
\f{(1+i)(\sqrt{2}+1)}{\sqrt{2}}
=
\big{(}{\overline{\cY^{(0)}}}\big{)}^{-1}
\,,
\ee
which would correspond to an opposite triangle angle of $3\pi/4$ and whose geometric interpretation is not quite clear.

Since the cube and the double pyramid are dual to each other, it is natural to consider the cubic graph and its dual double pyramid geometry. The regular double pyramid is a straightforward 2d triangulation, obtained by gluing together two equilateral pyramids, thus both with height $h=\sqrt{2}$, in which case all the triangle angles are equal to $\pi/3$ and all the dihedral angles are equal to $\textrm{arccos}\,1/3$. The geometric ansatz for Ising zeros is homogeneous with all couplings equal to:
\be
Y^{(0)}=\f13(\sqrt{2}+i)\,.
\ee
The cubic graph's loop polynomial for homogeneous couplings yields:
\be
P_{cube}[Y]=1+6Y^{4}+16Y^{6}+9Y^{8}
\,,
\ee
whose roots are the trivial $\pm i$ and then $\pm Y^{(0)}$ and its complex conjugate $\pm\overline{Y^{(0)}}$.

This setting allows a simple non-trivial check of the high T/low T duality between the regular cube and equilateral double pyramid:
\be
\cD[Y^{(0)}]=\overline{\cY^{(0)}}
\,,\qquad
\cD[\overline{Y^{(0)}}]=\cY^{(0)}
\,,\qquad
\cD[-Y^{(0)}]=\cY^{(1)}
\,,\qquad
\cD[-\overline{Y^{(0)}}]=\overline{\cY^{(1)}}
\,.
\ee
Now we would like to move on to a more intricate analysis of the double pyramid configuration, which allows to probe both convex and concave geometrical configurations and thus provide a further test of the geometrical formula, especially of the prescription for the sign of the dihedral angles.

\section{The flexible double pyramid}
\label{sec:doubleP}

\begin{figure}[h!]

\centering
\begin{tikzpicture}[scale=.7]

\coordinate(a) at (0,0) ;
\coordinate(b) at (0,2);
\coordinate(c) at (-1,-1.73);
\coordinate(d) at (-1,.27);
\coordinate(e) at (2,-.2);
\coordinate(f) at (-3,-.2);

\coordinate(x) at (-1.5,-2.595) ;
\coordinate(z) at (0,3) ;
\coordinate(y) at (3,0) ;

\draw (a)++(.1,-.3) node{$A$};
\draw (b)++(.3,.2)node{$B$};
\draw (c) ++(.2,-.2)node{$D$};
\draw (d) ++(-.2,.3)node{$C$};
\draw (e) ++(.2,0)node{$E$};
\draw (f) ++(-.2,0)node{$F$};


\draw[very thick] (a)--(b)--(d)--(c)--(a);
\draw[very thick, dashed] (e)--(a)--(f);
\draw[very thick] (e)--(b)--(f);
\draw[very thick] (e)--(c)--(f);
\draw[very thick] (e)--(d)--(f);
\draw[->,>=stealth] (a)--(x) node[left]{$x$};
\draw[->,>=stealth] (a)--(y)node[above]{$y$};
\draw[->,>=stealth] (a)--(z)node[right]{$z$};




\coordinate(A) at (8,0) ;
\coordinate(B) at (8,2);
\coordinate(C) at (7,-1.73);
\coordinate(D) at (7,.27);
\coordinate(E) at (10,2.5);
\coordinate(F) at (5,2.5);

\coordinate(X) at (6.5,-2.595);
\coordinate(Z) at (8,3);
\coordinate(Y) at (11,0);

\draw (A)++(-.4,-0.1) node{$A$};
\draw (B)++(-.2,.3)node{$B$};
\draw (C) ++(.2,-.2)node{$D$};
\draw (D) ++(-.3,-.2)node{$C$};
\draw (E) ++(.2,0)node{$E$};
\draw (F) ++(-.2,0)node{$F$};

\draw[very thick] (A)--(B)--(D)--(C)--(A);
\draw[very thick, dashed] (E)--(A)--(F);
\draw[very thick] (E)--(B)--(F);
\draw[very thick] (E)--(C)--(F);
\draw[very thick] (E)--(D)--(F);
\draw[->,>=stealth] (A)--(X)node[left]{$x$};
\draw[->,>=stealth] (A)--(Y)node[above]{$y$};
\draw[->,>=stealth] (A)--(Z)node[right]{$z$};


\coordinate(BCE) at (18,2.5) ;
\coordinate(ABE) at (17.5,-.5) ;
\coordinate(CDE) at (18.5,.5) ;
\coordinate(ADE) at (18,-2.5) ;

\draw[blue] (BCE) node {$\bullet$} node[above right] {$BCE$} ;
\draw[blue] (ABE) node {$\bullet$}  node[right] {$ABE$} ;
\draw[blue] (CDE) node {$\bullet$}  node[ right] {$CDE$} ;
\draw[blue] (ADE) node {$\bullet$}  node[right] {$ADE$} ;

\draw[purple, thick] (BCE)--node[above left]{$Y_{su}$}(ABE)--node[left]{$Y_{sd}$}(ADE)--node[below right]{$Y_{sd}$}(CDE)--node[right]{$Y_{su}$}(BCE);

\coordinate(BCF) at (14.5,2.5) ;
\coordinate(ABF) at (15,-.5) ;
\coordinate(CDF) at (14,.5) ;
\coordinate(ADF) at (14.5,-2.5) ;

\draw[blue] (BCF) node {$\bullet$}  node[above left] {$BCF$} ;
\draw[blue] (ABF) node {$\bullet$} node[left] {$ABF$} ;
\draw[blue] (CDF) node {$\bullet$} node[ left] {$CDF$} ;
\draw[blue] (ADF) node {$\bullet$} node[left] {$ADF$} ;

\draw[purple, thick] (BCF)--node[above right]{$Y_{su}$}(ABF)--node[right]{$Y_{sd}$}(ADF)--node[below left]{$Y_{sd}$}(CDF)--node[left]{$Y_{su}$}(BCF);

\draw[purple, thick] (BCE)--node[above]{$Y_{u}$} (BCF);
\draw[purple, thick] (ABE)--node[below]{$Y_{s}$}(ABF);
\draw[purple, thick] (CDE)--node[above]{$Y_{s}$}(CDF);
\draw[purple, thick] (ADE)--node[below]{$Y_{d}$}(ADF);

\end{tikzpicture}

\caption{
Double pyramid and its dual graph. The two mirror pyramids, with summits respectively $E$ and $F$, are glued by their common square base $(ABCD)$. We have drawn two examples of configuration with different positions for the summits $E$ and $F$: a convex configuration on the left, and a non-convex configuration on the right. Completely to the right, we have drawn the graph dual to the triangulation: the graph nodes in {\blue blue} are dual to the triangles and graph links in {\purple purple} are dual to the edges shared by neighbouring triangles. This graph naturally forms a cube. The Ising couplings $Y_{\ell}$ dress the graph links. Due to the symmetry of the geometry of our double pyramid, some of those couplings are assumed to be equal. We refer to them according to their position, {\bf u}p, {\bf d}own and {\bf s}ide.
\label{fig:doubleP}}

\end{figure}
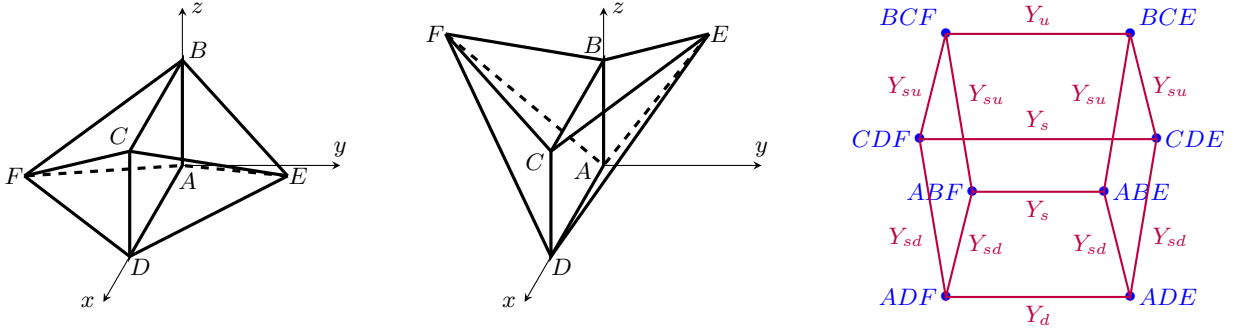

Let us consider the square-base symmetric double pyramid triangulation, made of two mirror-image sets of four triangles glued together, as drawn on fig.\ref{fig:doubleP}.
So we have four points forming a square, $(0,0,0)$, $(0,0,2)$, $(2,0,2)$, $(2,0,0)$, and back to the origin, plus the two mirror-image summits $(1,h,z)$ and $(1,-h,z)$ with the pyramid height $h>0$ and the summit drift (or ``tilt") $z\in\R$. An edge is drawn from each summit to the four points of the square, thus forming two square-base pyramids glued together.
When $z\in[0,2]$, this double pyramid is convex. When $z$ is larger than 2, the two upper triangles form a concavity. And when $z$ is negative, $z<0$, the two lower triangles form a concavity. Below, we will assume that $z$ is always positive, $z\ge 0$, and take special care of distinguishing the convex case $0\le z \le 2$ from the non-convex case $z\ge 2$.

The dual graph has one node dual to each triangle, thus two sets of 4 nodes. Each set is dual to its respective pyramid: they form a 4-cycle, which we can draw as a square, around the four triangles of the pyramid. Then each node is linked to its mirror image. Those 8 nodes, with 12 links, actually draw a cube, as illustrated on fig.\ref{fig:doubleP}. A careful counting of the cycles on the cube yields its loop polynomial, which reads after factorization:
\beq
P_{cube}[Y_{u},Y_{d},Y_{s},Y_{su},Y_{sd}]
&=&
Y_{s}^{2}(Y_{sd}^{2}+Y_{su}^{2})^{2}
+
(1+Y_{sd}^{2}Y_{su}^{2})^{2}
+
2Y_{s}Y_{u}Y_{su}^2(1+Y_{sd}^{2})^{2}
+
2Y_{s}Y_{d}Y_{sd}^2(1+Y_{su}^{2})^{2}
\nn\\
&&
+
4Y_{u}Y_{d}Y_{su}^{2}Y_{sd}^{2}
+
4Y_{u}Y_{d}Y_{s}^{2}Y_{su}^{2}Y_{sd}^{2}
\,.
\eeq

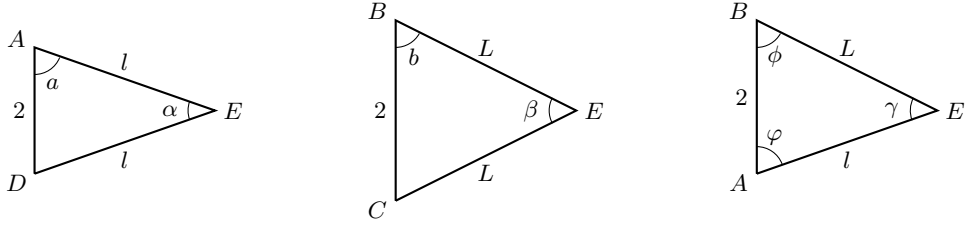
\begin{figure}[h!]

\centering
\begin{tikzpicture}[scale=1.2]

\draw (3.3,.8) node{$A$};
\draw (3.3,-.8) node{$D$};
\draw (5.7,0) node{$E$};

\coordinate(A) at (3.5,.7) ;
\coordinate(D) at (3.5,-.7);
\coordinate(E1) at (5.5,0);
\centerarc[](E1)(160:200:.3);
\draw (5,0) node{$\alpha$};
\centerarc[](A)(270:338:.3);
\draw (3.7,.3) node{$a$};

\draw[thick] (D)-- node[below]{$l$}(E1)-- node[above]{$l$}(A)--node[left]{$2$}(D);

\draw (7.3,1.1) node{$B$};
\draw (7.3,-1.1) node{$C$};
\draw (9.7,0) node{$E$};

\coordinate(B) at (7.5,1) ;
\coordinate(C) at (7.5,-1);
\coordinate(E2) at (9.5,0);
\centerarc[](E2)(152:208:.3);
\draw (9,0) node{$\beta$};
\centerarc[](B)(270:331:.3);
\draw (7.7,.6) node{$b$};

\draw[thick] (C)-- node[below]{$L$}(E2)-- node[above]{$L$}(B)--node[left]{$2$}(C);

\draw (11.3,1.1) node{$B$};
\draw (11.3,-.8) node{$A$};
\draw (13.7,0) node{$E$};

\coordinate(C2) at (11.5,1) ;
\coordinate(D2) at (11.5,-.7);
\coordinate(E3) at (13.5,0);
\centerarc[](E3)(152:201:.3);
\draw (13,0) node{$\gamma$};
\centerarc[](C2)(270:331:.3);
\draw (11.7,.6) node{$\phi$};
\centerarc[](D2)(21:90:.3);
\draw (11.7,-.3) node{$\vphi$};

\draw[thick] (D2)-- node[below]{$l$}(E3)-- node[above]{$L$}(C2)--node[left]{$2$}(D2);

\end{tikzpicture}

\caption{
Triangle angles in the double pyramid.
\label{fig:pyramidangles}}

\end{figure}

Looking at the 2d geometry, there are three different triangles: the upper one $(BCE)$, the lower one $(ADE)$ and the side one $(ABE)$. All other triangles are identical to those ones or mirror images. Knowing the length of upper triangles edges and of the lower triangles edges in terms of the free coordinates, $h$ and $z$,
\be
l=
L_{AE}=
L_{DE}=
L_{AF}=
L_{DF}=
\sqrt{1+h^{2}+z^{2}}
\,,\qquad
L=
L_{BE}=
L_{CE}=
L_{BF}=
L_{CF}=
\sqrt{1+h^{2}+(2-z)^{2}}
\,,
\ee
\be
\text{with inverse formulas:}\qquad
z
=
\f{1}{4}
(l^{2}-L^{2}+4)
\,,\qquad
1+h^{2}
=
\f{1}{4^{2}}(l-L+2)(L+l+2)(L+l-2)(L-l+2)
\,,
\ee
we can compute all the triangle angles, defined in fig.\ref{fig:pyramidangles}. More precisely we give the half-angle tangents, using directly their formula in terms of edge lengths:
\be
\tan^{2}\f\alpha2=\f1{l^{2}-1}
\,,\quad
\tan^{2}\f a2=\f{l-1}{l+1}
\,,\qquad
\tan^{2}\f\beta2=\f1{L^{2}-1}
\,,\quad
\tan^{2}\f b2=\f{L-1}{L+1}
\,,
\ee
\be
\tan^{2}\f\gamma2=
\f{4-(l-L)^{2}}{(l+L)^{2}-4}
\,,\qquad
\tan^{2}\f\vphi2=
\f{L^{2}-(l-2)^{2}}{(l+2)^{2}-L^{2}}
\,,\qquad
\tan^{2}\f\phi2=
\f{l^{2}-(L-2)^{2}}{(L+2)^{2}-l^{2}}
\,,
\ee
which satisfy in particular the law of half-tangents:
\be
\tan\f\gamma2\tan\f\vphi2+\tan\f\vphi2\tan\f\phi2+\tan\f\phi2\tan\f\gamma2
=1
\,.
\ee
Let us keep in mind that, since all the triangle angles are in $[0,\pi]$, all the half-angle tangents are positive.

As for the  dihedral angles, we compute them using the scalar product between the outward oriented triangle normal vectors:
\be
\cos\f{\theta_{BC}}2
=\f h{\sqrt{L^{2}-1}}
\,,\quad
\sin\f{\theta_{BC}}2
=\f {2-z}{\sqrt{L^{2}-1}}
\,,\qquad
\cos\f{\theta_{AD}}2
=\f h{\sqrt{l^{2}-1}}
\,,\quad
\sin\f{\theta_{AD}}2
=\f {z}{\sqrt{l^{2}-1}}
\,,
\ee
where the sign of the sines appropriately takes into account the sign convention described previously: we take a positive dihedral angle, $\theta\in[0,\pi]$, for a convex configuration, and a negative dihedral angle for an inward fold.
As for the other edges, we have:
\be
\cos\f{\theta_{AB}}2
=\f h{\sqrt{1+h^{2}}}
\,,\quad
\sin\f{\theta_{AB}}2
=\f 1{\sqrt{1+h^{2}}}
\,,
\ee
\be
\cos\theta_{AE}
=
\f z{\sqrt{1+h^{2}}\sqrt{l^{2}-1}}
\,,\qquad
\cos\theta_{BE}
=
\f{(2-z)}{\sqrt{1+h^{2}}\sqrt{L^{2}-1}}
\,.
\ee
This allows to compute the ansatz for geometric Ising zeros:
\be
Y_{u}^{(0)}
=
e^{i\f{\theta_{BC}}2}\tan\f\beta2
=
\f{h+i(2-z)}{(L^{2}-1)}
\,,\qquad
Y_{d}^{(0)}
=
e^{i\f{\theta_{AD}}2}\tan\f\alpha2
=
\f{h+iz}{(l^{2}-1)}
\,,
\ee
\be
Y_{s}^{(0)}
=
e^{i\f{\theta_{AB}}2}\tan\f\gamma2
=
\f{h+i}{\sqrt{1+h^{2}}}\,
\sqrt{\f{(l-L+2)(L-l+2)}{(L+l+2)(L+l-2)}}
=
\f{4(h+i)}{(L+l+2)(L+l-2)}
\,,
\ee
\be
(Y_{su}^{(0)})^2
=
e^{i{\theta_{BE}}}\,{\tan\f\vphi2\tan\f b2}
=
\f{4\,(2-z+ihL)}{(L+1)(l+L+2)(l-L+2)}
\,,
\ee
\be
(Y_{sd}^{(0)})^2
=
e^{i{\theta_{AE}}}\,{\tan\f\phi2\tan\f a2}
=
\f{4\,(z+ihl)}{(l+1)(L+l+2)(L-l+2)}
\,.
\ee
%
%
Plugging these formulas in the loop polynomial, it is direct to check numerically that   our $Y$-ansatz, in terms of $l$ and $L$, or equivalently $h$ and $z$, are actually roots of the loop polynomial $P_{cube}$, in both convex and non-convex cases. It is also straightforward, though tedious and lengthy, to show this explicitly analytically by setting all the terms on a common denominator.
We do not write the intermediate calculations, because they are lengthy and do not seem enlightening, and can also be swiftly checked on a formal calculator such as Mathematica:
\be
\forall h,\,z,\qquad P_{cube}
\Big{[}Y_{u}^{(0)},Y_{d}^{(0)},Y_{s}^{(0)},Y_{su}^{(0)},Y_{sd}^{(0)}\Big{]}
=0
\,.
\ee
This provides a non-trivial analytical check of the geometrical formula for Ising zeros, for a continuous set of variable parameters and allowing for both convex and concave geometries, going beyond the tetrahedron geometry.


\section{Numerical Check on Random Polyhedra}
%

We have tested explicitly the validity of the geometric formula \eqref{eq_Ising_zeroes} for Ising zeroes on finite planar graphs for simple configurations: the double triangle, the tetrahedron, the pyramid, cube and double pyramid.
In this section, we propose a more extensive validation by performing a thorough numerical  check of the formula for random  triangulations on the sphere.
To this purpose, we introduce a method to generate random convex polyhedra and adapt it to also generate random non-convex closed ones. 
We then check that the geometric couplings are indeed zeros of the loop polynomial for every random generated triangulation and study the growth of its modulus
in a neighborhood of those critical couplings.

A side-product of our analysis is a method to detect the convexity or concavity of the gluing of adjacent triangles  around each edge of a closed polyhedron.

\subsection{Numerical Method}
\label{sec_nummethod}

\begin{figure}[htb]
             \begin{tabular}{|c|M{38mm}M{38mm}M{38mm}M{38mm}|}
        \hline        & 14 vertices & 15 vertices & 16 vertices & 17 vertices\\
         & (24 faces) & (26 faces) & (28 faces) & (30 faces)\\
        \hline
                \begin{tabular}{c} On $\cS^2$ \end{tabular}
              & \begin{tabular}{c}    \includegraphics[width=2.5cm]{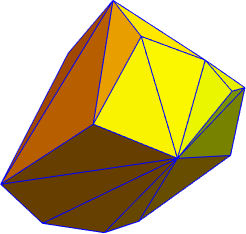}\end{tabular}&\begin{tabular}{c} \includegraphics[width=2.5cm]{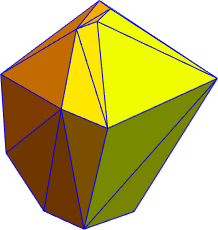}\end{tabular}&
\begin{minipage}{2.5cm}        
\vspace*{4mm}        
        \begin{tabular}{c}  \includegraphics[width=2.5cm]{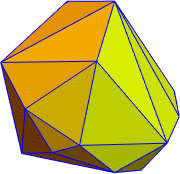}\end{tabular}\vspace*{4mm} 
        \end{minipage}&\begin{tabular}{c}  \includegraphics[width=2.5cm]{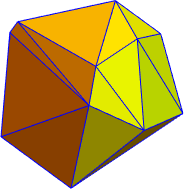}\end{tabular}\\
        \hline
        \begin{tabular}{c} Rescaled \end{tabular}&\begin{tabular}{c}    \includegraphics[width=2.5cm]{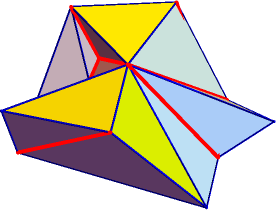}\end{tabular}&\begin{tabular}{c} \includegraphics[width=2.5cm]{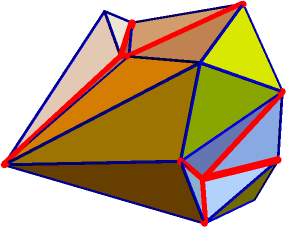}\end{tabular}& \begin{minipage}{2.5cm}        
\vspace*{4mm} \begin{tabular}{c} \includegraphics[width=2.5cm]{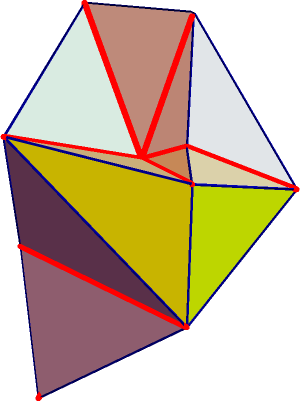}
       \end{tabular}\vspace*{4mm} \end{minipage}&\begin{tabular}{c}  \includegraphics[width=2.5cm]{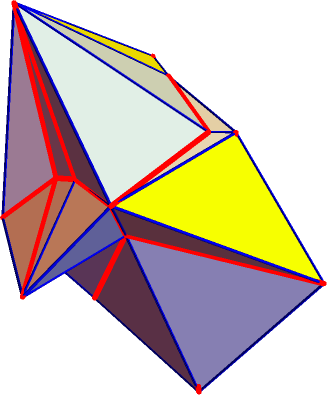}\end{tabular}\\
       \hline
    \end{tabular}
  \caption{Polyhedra for 14, 15, 16 and 17 vertices are represented. The first row shows  convex polyhedra corresponding to  Delaunay triangulations on the unit 2-sphere $\cS^2$, while the second row shows their locally-rescaled non-convex versions. The edges corresponding to non-convex angles are plot in red.}\label{fig_polyhedra}
\end{figure}

We have numerically computed the loop polynomial defined in  \eqref{eq:partition_looppol}, by brutally summing over all spins $\sigma_{n}=\pm$, associated to graphs dual to random polyhedra  up to 30 faces (so that each graph node corresponds to a face of the polyhedron,
as drawn on fig.\ref{fig_polyhedra}).
We have systematically checked its value on the geometric couplings given  by equation \eqref{eq_Ising_zeroes}, thereby confirming that they are indeed zeros of the  Ising partition function.

Our method to generate random convex and non-convex  triangulations, with spherical topology, proceeds in two steps: first generate Delaunay triangulations on the 2-sphere for random point distributions, which produces convex triangulations, and then locally rescale in a random way the radial coordinate for each point in order to allow for non-convex configurations.

In more details, our starting point is a random distribution\footnotemark{} of seeds on the unit 2-sphere $\cS^{2}$. These seeds will be the vertices of the triangulations. Then we associate a Delaunay triangulation, which draws triangles between those seeds and forms a convex polyhedron, as illustrated on fig.\ref{fig_polyhedra}.
In order to generate the Delaunay triangulation we have used standard algorithms \cite{Renka}. The output of the Delaunay generator returns the labels of the vertices belonging to each face of the associated convex polyhedron. 
Delaunay triangulations, and their dual Vorono\"i diagrams, are very useful and efficient mathematical concepts, allowing for an a posteriori identification of nearest neighbors for arbitrary point distributions, used in many fields of physics, for instance in image processing (sampling, compression and generation), statistical physics (e.g. for particle marking in hydrodynamics), quantum gravity (e.g. \cite{Gwynne:2018tau,Diaz-Polo:2013hga}), or even quantum information (e.g. \cite{Nielsen2008,kato2006,PhysRevB.100.155402}).
Although the notion of nearest neighbor is not directly relevant to our analysis, using Delaunay triangulations automatically ensures  a clean identification of a convex triangulation on the 2-sphere, with non-overlapping faces, and its dual graph.

The next step is to perform a local rescaling of each vertex in the radial direction, that is, we multiply the coordinates of each vertex by a random factor, here chosen in the range $[1,4]$. The combinatorial structure of the faces of the triangulation remains unaltered, but this rescaling procedure  changes the folds between the faces and allow for non-convex configurations, as shown on fig.\ref{fig_polyhedra}, thereby allowing us to test the geometric formula for Ising zeros in general convex and non-convex configurations.
Notice that, depending on the specific case, the rescaling may be too big and the structure of closed polyhedron could be lost. These cases have been excluded.
\footnotetext{
In fact we have mainly worked with uniform distributions of seeds on the 2-sphere. Although we have also successfully checked the results with other distributions (for example distributions with higher number of seeds near the poles of the sphere), the final effect on the shape of the polyhedron of the initial distribution used is not as important as the effect of the rescaling of each vertex  to get non-convex polyhedra.
} 

In order to compute the Ising zeros associated to the  polyhedron, given by equation (\ref{eq_Ising_zeroes}), we need to compute the planar and dihedral angles associated to each edge. Given that we have control over the vertices that define each edge and each face of the polyhedron, the computation of the planar angles is straightforward (computing the scalar product of the edges, for instance). Nevertheless, in order to compute the exterior dihedral angles, we need to compute the outward-oriented normal vector to each face. Computing a normal vector is straightforward, but the choice of the outward direction may be highly non-trivial in the general case for non-convex polyhedra. 

Given that the surface of the polyhedron has the topology of a 2-sphere, we may determine the orientation with a cyclic ordering of the vertices of each face, as explained in section \ref{sec:orientation}. However, from a computational point of view, assigning a proper cyclic order to each face is not easy.
In our construction, the non-convex polyhedron is obtained as a deformation of a convex one that has its vertices on the unit sphere centred at the origin. So we can determine the outward orientation for the initial convex polyhedron before rescaling and retain the same orientation for the actual non-convex  polyhedron after rescaling.
Indeed, the choice of the outward normal is easily made for the convex polyhedron by checking if the normal vector computed with the coordinates of the vertices of the face has positive projection over the position vector of any of the vertices of the face. Once the outward-oriented normal is chosen, the positive orientation of the face is determined, i.e., the cyclic order of each triangle, and this orientation will be preserved for the rescaled non-convex polyhedron.

\begin{figure}[htb!]
\vspace*{.9cm}
\begin{overpic}[abs,unit=1mm,width=7cm
]{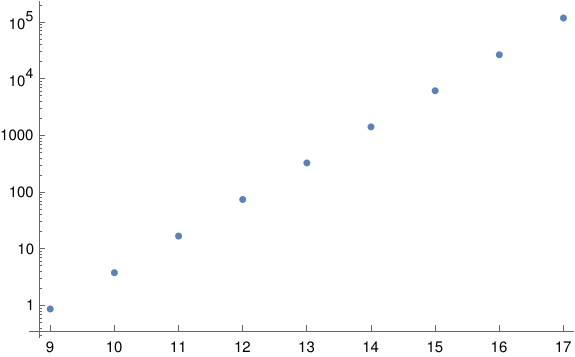}
\put(-2,48){computation}
\put(0,45){time (s)}
\put(71,2){\parbox{2cm}{number of\\ vertices}}
\end{overpic}
\caption{The plot represents the computation time (in logarithmic scale) of the loop polynomial for dual polyhedra with increasing number of vertices (or faces). It is clear the linear behavior of the graph, that implies exponential growth of the time.}\label{fig:time}
\end{figure}

Finally, we compute the loop polynomial \eqref{eq:partition_looppol} on the graph dual to the triangulation: each face corresponds to a graph node and each pair of adjacent faces corresponds to a  link between those two graph nodes.
We need to sum over all the possible configurations of the values of spins $\sigma_n=\pm 1$ equivalently associated to the triangulation faces or to the graph nodes. Therefore, we need to generate all possible lists of length the number of faces with all the configurations of $\pm 1$. That is a total of $2^N$ configurations, where $N$ is the total number of faces. Hence, the computational cost for large number of vertices (faces) is significant. In fact, the computation time grows exponentially with the number of vertices (see fig.\ref{fig:time}).
 Although we are aware that our computation may possibly be optimized, for our interests in this paper it is enough with the 30 faces that we have been able to achieve using Mathematica on a 
single CPU\footnotemark{}.
\footnotetext{We have used a single CPU but with 768 GB of RAM memory. Large RAM memory is necessary especially for the cases with more than 12 vertices (20 faces)}

\subsection{Checking the Geometric Formula for Ising Zeros}
\label{sec:numerics}

The implementation of the method presented in section \ref{sec_nummethod} in order to compute the Ising partition function associated to a 2d closed triangulation (both convex and non-convex) has been a success. More concretely, we have been able to check the expression for the zeroes of the partition function  \eqref{eq_Ising_zeroes} for a large number (hundreds) of different configurations of both convex and non-convex polyhedra.
In the following we will comment the main results (shared by each configuration we have tried), limitations of the code and consequences.

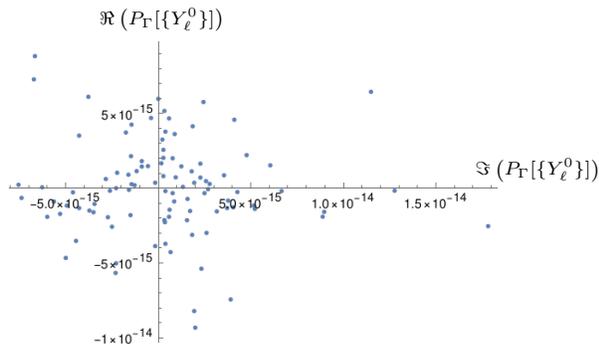
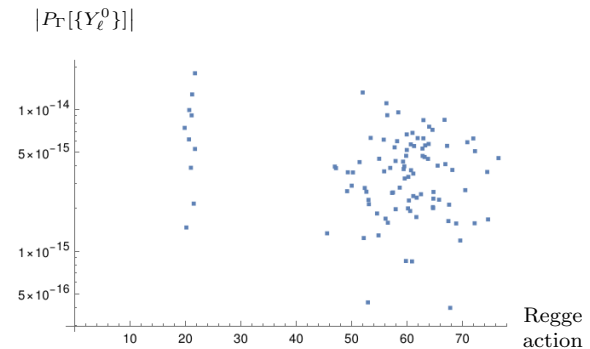
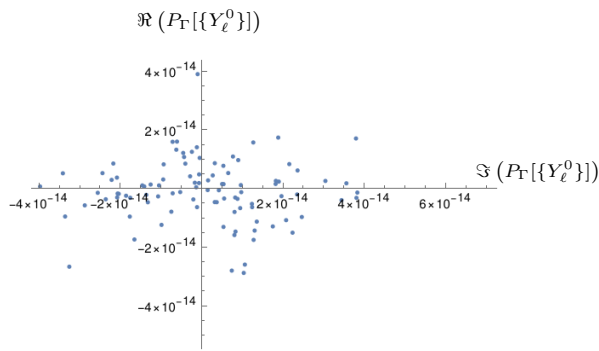
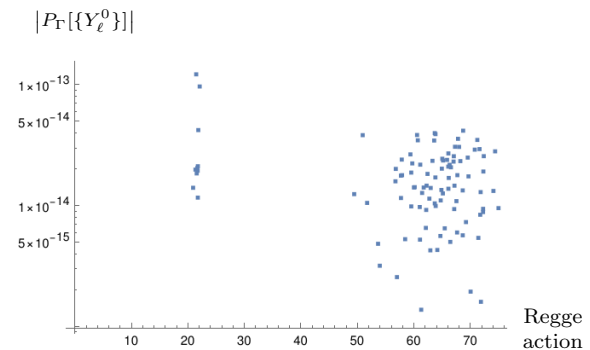
\begin{figure}[htb!]
\vspace*{.9cm}
\subfloat[Real versus imaginary part of the loop polynomial for polyhedra with 11 vertices. \label{fig:cloud1}]{
\hspace*{-1cm}
\begin{overpic}[abs,unit=1mm,width=6.5cm
]{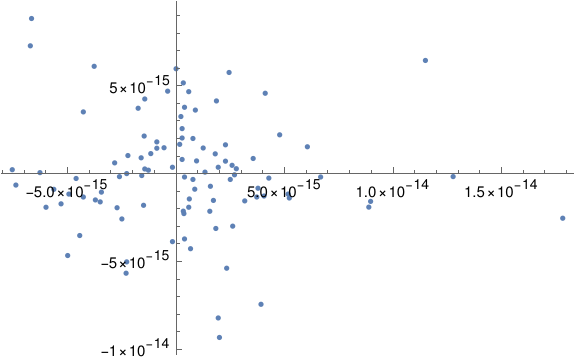}
\put(12,43){{\scriptsize $\Re\left(P_{\Gamma}[\{Y_{\ell}^0\}]\right)$}}
\put(62,23){\scriptsize $\Im\left(P_{\Gamma}[\{Y_{\ell}^0\}]\right)$}
\end{overpic}
}
\hspace{22mm}
\subfloat[Absolute value of the loop polynomial versus the value of the Regge action for polyhedra with 11 vertices.\label{fig:cloud2}]{
\begin{overpic}[abs,unit=1mm,width=6.5cm
]{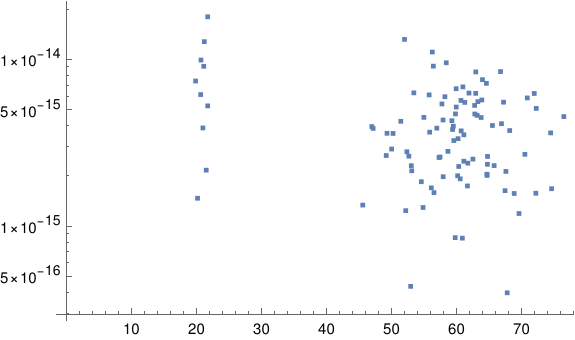}
\put(2,43){{\scriptsize $\big|P_{\Gamma}[\{Y_{\ell}^0\}]\big|$}}
\put(61,2){\parbox{2cm}{\footnotesize Regge\\ action}}
\end{overpic}
}\\[1cm]
\subfloat[Real versus imaginary part of the loop polynomial for polyhedra with 13 vertices. \label{fig:cloud3}]{
\hspace*{-1cm}
\begin{overpic}[abs,unit=1mm,width=6.5cm
]{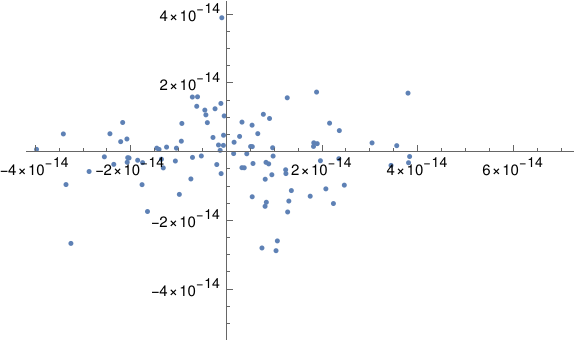}
\put(17,43){{\scriptsize $\Re\left(P_{\Gamma}[\{Y_{\ell}^0\}]\right)$}}
\put(62,23){\scriptsize $\Im\left(P_{\Gamma}[\{Y_{\ell}^0\}]\right)$}
\end{overpic}
}
\hspace{22mm}
\subfloat[Absolute value of the loop polynomial versus the value of the Regge action for polyhedra with 13 vertices.\label{fig:cloud4}]{
\begin{overpic}[abs,unit=1mm,width=6.5cm
]{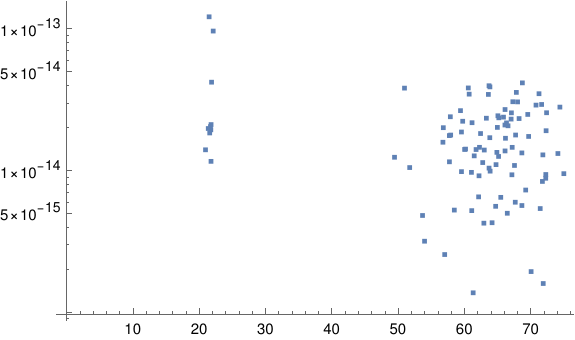}
\put(2,43){{\scriptsize $\big|P_{\Gamma}[\{Y_{\ell}^0\}]\big|$}}
\put(61,2){\parbox{2cm}{\footnotesize Regge\\ action}}
\end{overpic}
}
\caption{Cloud of trial points for 11 (fig.\ref{fig:cloud1} and fig.\ref{fig:cloud2}) and 13 vertices (fig.\ref{fig:cloud3} and fig.\ref{fig:cloud4}).
In each plot the results for 100 different configurations are shown. More concretely, we have considered 10 different initial distribution of seeds on the unit sphere (whose Delaunay triangulation gives a convex polyhedron) and for each of these convex configurations we have considered 9 different local rescalings corresponding to non-convex polyhedra.
 In the first plot of each row (fig.\ref{fig:cloud1} and fig.\ref{fig:cloud3}) it is represented the real part versus the imaginary part of the evaluation of the loop polynomial. On fig.\ref{fig:cloud2} and fig.\ref{fig:cloud4}) the modulus of the loop polynomial versus the extrinsic curvature integral is represented. The vertical line of dots around a value of 20 in the Regge action corresponds to the 10 convex configurations we considered. This behavior of the value of the Regge action for convex configurations requires deeper study but, in any case, it does not affect to our results.
 Finally, comparing plots \ref{fig:cloud2} and \ref{fig:cloud4} we observe that the results for the zeroes of the loop polynomial (i.e., Ising partition function) are statistically better for the case of 11 vertices. This effect of the loss of accuracy (due to numerical error) with increasing number of vertices is also shown in fig.\ref{fig_precission}.
  }\label{fig:cloud}
\end{figure}

The main result is that our numerical check for the formula \eqref{eq_Ising_zeroes} is highly satisfactory.
We computed the value of the  loop polynomial for one hundred different configurations of 11 vertices and another one hundred with 13 vertices, which we plotted on fig.\ref{fig:cloud}, as clouds of points. The plot of the real part versus the imaginary part clearly shows that both values are very small, of the order of magnitude of the numerical precision of our Mathematica, around $10^{-14}$, as expected. The plot  of the absolute value versus the value of the Regge action  allows to discriminate the various poyhedra with respect to their geometry and embedding in the 3d space, since the Regge action $\sum_{e}\theta_{e}L_{e}$ is a discretization of the integral of the extrinsic curvature of the surface as the sum of the edge lengths times the corresponding dihedral angles.

\begin{figure}[htb!]
\vspace*{.9cm}
\begin{overpic}[abs,unit=1mm,width=7cm
]{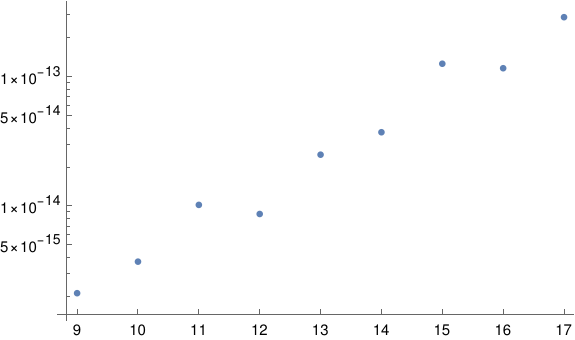}
\put(-2,45){$\big|P_{\Gamma}[\{Y_{\ell}^0\}]\big|$}
\put(71,2){\parbox{2cm}{number of\\ vertices}}
\end{overpic}
\caption{The plot shows the accuracy obtained by the numerical method for increasing number of vertices. In the horizontal axis the number of vertices of the triangulation is represented and in the vertical one the value of absolute value of the computed loop polynomial in logarithmic scale.
 Although the results are still very good, the plot shows that the numerical error grows exponentially with the number of vertices. Therefore, the code should be optimized in order to attempt the computation of large number of vertices.}\label{fig_precission}
\end{figure}

Furthermore, we have  run the code for different configurations of polyhedra up to 17 vertices, corresponding to 30 faces, and the results are always extremely close to zero (not zero because of numerical error), although the accuracy of the result gets exponentially worse  when increasing the number of vertices of the triangulation (see fig.\ref{fig_precission}). Nevertheless, for 17 vertices we are still obtaining errors of order $3\times 10^{-13}$. In any case, the examples of 17 vertices are also, due to computational costs (fig. \ref{fig:time}), the limit cases we are able to compute with our code and resources. Therefore, in order to be able to discuss examples with large number of vertices or faces (relevant in order to study the thermodynamic limit in general cases), it would be necessary to improve and optimize the method and the code to compute the Ising partition function.

\begin{figure}
\hspace*{-2cm}
\begin{minipage}{4cm}
\vspace*{1.5cm}
\includegraphics[height=3cm]{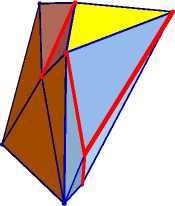}\vspace*{2.5cm}
\end{minipage}\hspace*{2cm}
\begin{minipage}{7cm}
\begin{overpic}[abs,unit=1mm,height=5cm
]{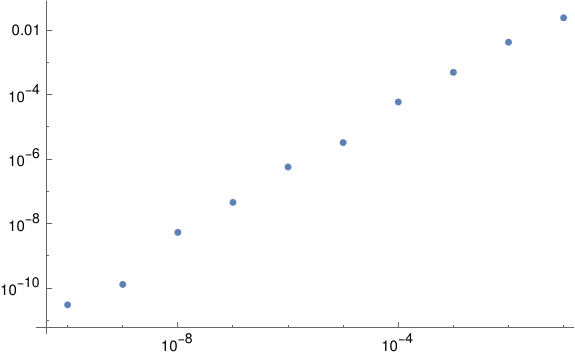}
\put(0,53){$\big|P_{\Gamma}[\{Y_{\ell}^0\}]\big|$}
\put(82,3){pert.}
\end{overpic}
\end{minipage}
\caption{We consider a non-convex polyhedron corresponding to a triangulation with 9 vertices. We plot with logarithmic scale in both axes the absolute value of the loop polynomial for couplings that are further and further from the critical coupling.}\label{fig_pert_allsigns}
\end{figure}

We have also checked that the zeroes given by the equation \eqref{eq_Ising_zeroes} are non-trivial. We have computed the loop polynomial for couplings $Y_{\ell}$ slightly perturbed around the critical couplings $Y_{\ell}^{(0)}$. More specifically, we have generated random perturbations $p_{\ell}$ of certain amplitude $a$ around the critical coupling. That is, 
\be
Y_{\ell}=Y_{\ell}^{(0)}+p_{\ell},\quad -a\leq p_{\ell} \leq a\quad \forall\, \ell\,.
\ee
Repeating the procedure for larger and larger values of the amplitudes, the results indicate that the absolute values of the loop polynomial increases in a linear way with the value of the amplitude of the perturbations, see fig.\ref{fig_pert_allsigns}.

\subsection{About non-convexity: testing the sign of the dihedral angles}

Once the formula \eqref{eq_Ising_zeroes} is checked, we may use it in order to study which of the edges of a close polyhedron (determined only by the vertices of the triangulation) are non-convex. 

Based on the characterization of the convexity-concavity of each edge of the polyhedron by the sign of the dihedral angle associated to it (see section \ref{sec:orientation}), the idea is to consider all the possible configurations of signs for the dihedral angles, i.e., a total of $2^E$ configurations, where $E$ is the number of edges of the polyhedron. 
We compute the associated loop polynomial for each of them and we select the two that cancel the loop polynomial (up to numerical error). There are always two of such configurations that differ on a global sign. This is, precisely, the overall sign that appears in equation \eqref{eq_Ising_zeroes} and that corresponds to consider either the interior or the exterior part of the polyhedron.

\begin{figure}[h!]
\hspace*{-5.8cm}
\begin{minipage}{4cm}
\vspace*{3cm}
\includegraphics[height=3cm]{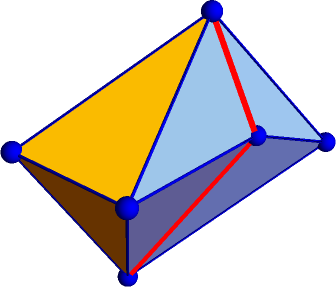}\vspace*{2.5cm}
\end{minipage}\hspace*{1cm}
\begin{minipage}{7cm}
\begin{overpic}[abs,unit=1mm,height=60mm
]{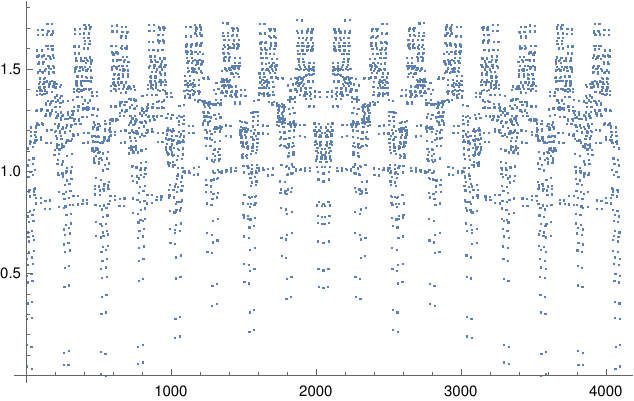}
\put(-7,63){$\big|P_{\Gamma}[\{Y_{\ell}[\{\theta_{\ell}\}]\}]\big|$}
\put(95,2){\parbox{2cm}{config.\\ number}}
\end{overpic}
\end{minipage}
\caption{Computation of the loop polynomial for each configuration of signs of the dihedral angles. For a triangulation of 6 vertices, we have 12 edges so we had to evaluate the loop polynomial $2^{12}$ times. Among all the possible results there are only 2 of them with the same minimum value of the loop polynomial. These two configurations of the signs of the dihedral angles differ only on a global sign that does not affect to the result.}\label{fig_allsigns}
\end{figure}

\begin{figure}[h!]
\hspace*{-5.8cm}
\begin{minipage}{4cm}
\vspace*{3cm}
\includegraphics[height=3cm]{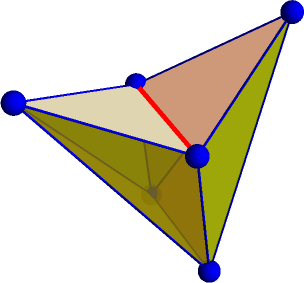}\vspace*{2.5cm}
\end{minipage}\hspace*{1cm}
\begin{minipage}{7cm}
\begin{overpic}[abs,unit=1mm,height=60mm
]{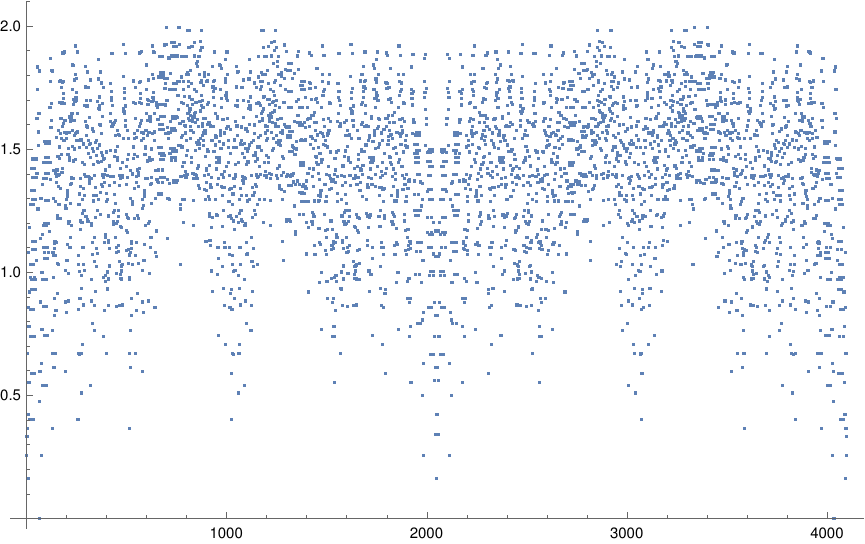}
\put(-7,63){$\big|P_{\Gamma}[\{Y_{\ell}[\{\theta_{\ell}\}]\}]\big|$}
\put(97,2){\parbox{2cm}{config.\\ number}}
\end{overpic}
\end{minipage}
\caption{Computation of the loop polynomial for each configuration of signs of the dihedral angles for the double pyramid, i.e, a total of $2^{12}$ configurations. In this particular case, among all the results the configuration with minimum loop polynomial is the number 65, corresponding to the following signs for the dihedral angles $(1, 1, 1, 1, 1, -1, 1, 1, 1, 1, 1, 1)$, and there exist another configuration with opposite signs.
Given that the negative dihedral angle corresponds to the sixth edge (represented in red), we have determined the convexity/concavity associated to each edge of the polyhedron. 
}\label{fig_allsigns_double}
\end{figure}

Due to computational time we have been able to implement this method only for polyhedra with low number of vertices. In fig.\ref{fig_allsigns} we show the result for a polyhedron of 6 vertices with two concave edges and in fig.\ref{fig_allsigns_double} we represent the result for the case of the flexible double pyramid studied in section \ref{sec:doubleP}.
Although this procedure is very expensive from a computational point of view, the result is non-trivial and general.


\section{Topology matters: the Donut pitfall}

The conjecture about the geometric formula for 2d Ising zeros assumes that the graph is planar and thus its dual triangulation has the topology of a 2-sphere. It is therefore natural to inquire whether this is a necessary restriction and to check how much topology does matter here. Indeed, the geometric formula comes from a duality between the 2d Ising model and 3d quantum gravity amplitudes \cite{Bonzom:2015ova} and the ensuing saddle point analysis of the semi-classical regime of the quantum gravity path integral \cite{Bonzom:2024zka}.
Although this duality has been proven only for a spherical topology and has not been extended beyond, we naturally expect that the non-trivial connections and transports around the non-contractible cycles would have a non-negligible impact on the semi-classical amplitudes, as explored for instance for toroidal topology in \cite{Dittrich:2017hnl,Dittrich:2017rvb,Dittrich:2018xuk,Livine:2021sbf},  and thus on the resulting geometric parametrization for Ising zeros.

\begin{figure}[ht!]
\centering
\subfloat[Prismatic torus. \label{fig:prismatic_donut}]{
\includegraphics[height=4.5cm]{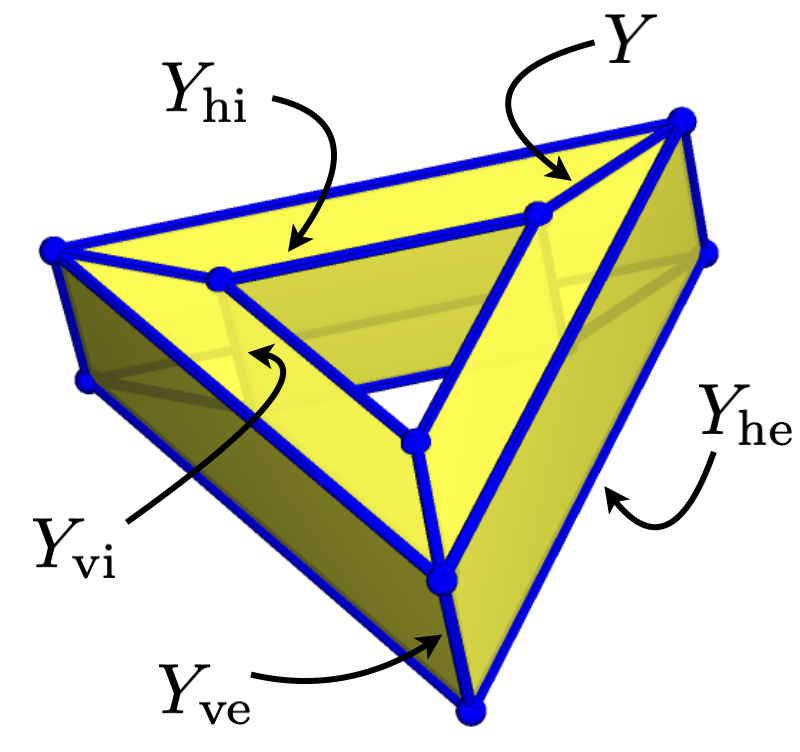}}
\hspace{1.6cm}
\subfloat[Wedgy  torus with isosceles corner triangle. \label{fig:isosceles_donut}]{
\includegraphics[height=3.8cm]{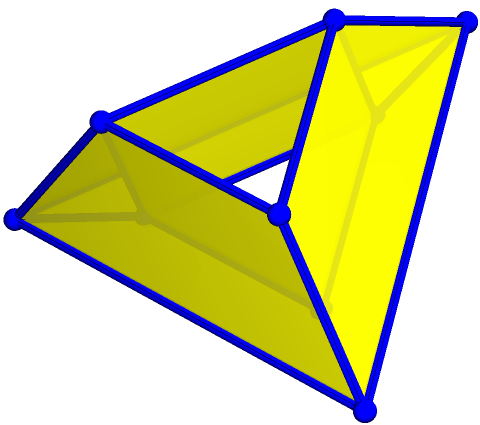}}
\hspace{1.6cm}
\subfloat[Wedgy  torus with equilateral corner triangle. \label{fig:equilateral_donut}]{
\includegraphics[height=3.8cm]{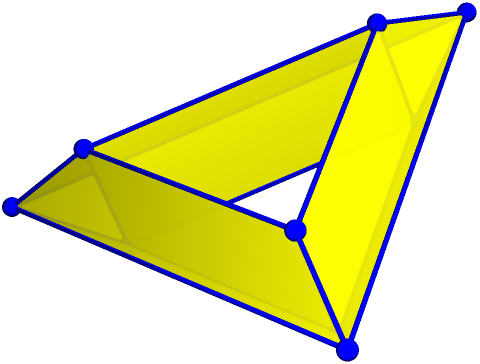}}
\caption{
We have considered three different configurations of graphs with a toroidal topology.
The corresponding geometries are made from gluing three blocks forming an inner ring. On the left, the prismatic torus (fig.\ref{fig:prismatic_donut}) is built from three prisms. There are five types of edges, with corresponding couplings $Y,Y_{\text{he}},Y_{\text{hi}},Y_{\text{ve}},Y_{\text{vi}}$, according to whether they are horizontal or vertical, and in the inner or outer rings. This configuration is rather symmetric due to the rectangular nature of the vertical faces, and its  loop polynomial may be explicitly written in a not-too-long fashion, see equation \eqref{eq:looppol_torus}.
The other two configurations which we considered (fig.\ref{fig:isosceles_donut} and fig.\ref{fig:equilateral_donut}) are built out of three wedges. The advantage is that there are less faces (no outer face) nevertheless makes the configuration less symmetric and the expression for the loop polynomial of those wedgy prisms is more intricate.
\label{fig_prismic_torus}
}
\end{figure}

Following this exploratory logic, we extend in this section the analysis carried out up to now, from configurations with the topology of the 2-sphere to the case of configurations with toroidal topology. We consider the donut configurations drawn in fig.\ref{fig_prismic_torus}. We proceed in an analogous way and  check if the proposed formula \eqref{eq_Ising_zeroes} for the zeros of the partition function \eqref{eq_partition1} is still valid for non-spherical topologies. Anticipating the result of the numerical analysis, we find that the formula is not valid anymore and that the 2d topology does actually matter.

We first consider the prismatic torus represented in fig.\ref{fig:prismatic_donut}, made of three prisms glued together to form a donut with inner and outer rings. The bottom and top are rings made of three isosceles trapezoids. Those two horizontal layers are linked together by vertical rectangular faces, forming the inner and outer rings. We choose the inner and outer limiting triangles to be equilateral, so that we can play with three parameters: the inner radius $r$, which is the distance between the center of the triangle to its summits so that the edge length is $l=r\sqrt{3}$, the outer radius $R$, and the height of the donut $h$.
Given the symmetries of this configuration, we will only have 5 different coupling constants in the loop polynomial: $Y$, corresponding to the edge between the inner and outer triangles, $Y_{\text{hi}}$ and  $Y_\text{he}$ corresponding to the horizontal edges of the inner and outer triangles respectively, and $Y_{\text{vi}}$ and  $Y_\text{ve}$ associated to the vertical edges of the inner and outer rectangles respectively.
The loop polynomial after factorization, and up to a global combinatorial factor of 2 power the number of faces, so $2^{12}$, reads:
\begin{align}
P_{\text{torus}}&[Y,Y_{\text{hi}},Y_\text{he},Y_\text{vi},Y_\text{ve}]=(Y+1)^2 \left(Y_{\text{ve}}+1\right) \left(Y_{\text{vi}}+1\right)
   \left(Y_{\text{he}}^6 Y_{\text{hi}}^2 \left(\left(Y_{\text{ve}}-1\right)
   Y_{\text{ve}}+1\right) \left(((Y-1) Y+1)^2 Y_{\text{hi}}^4
\right.\right.\nn
\\
   &\left.\left.   
   \left(\left(Y_{\text{vi}}-1\right) Y_{\text{vi}}+1\right)+6 Y ((Y-1) Y+1)
   Y_{\text{hi}}^2 Y_{\text{vi}}+3 Y^2
   \left(Y_{\text{vi}}^2+Y_{\text{vi}}+1\right)\right)+3 Y_{\text{he}}^4
   \left(2 Y \left(Y^2+Y+1\right) 
   \right.\right.\nn
\\
   &\left.\left. Y_{\text{hi}}^2 \left(Y_{\text{ve}}
   \left(Y_{\text{vi}}
   \left(Y_{\text{ve}}+Y_{\text{vi}}+2\right)+1\right)+Y_{\text{vi}}\right)+2 Y
   ((Y-1) Y+1) Y_{\text{hi}}^6 Y_{\text{ve}} \left(\left(Y_{\text{vi}}-1\right)
   Y_{\text{vi}}+1\right)
\right.\right.\nn\\   
   &\left.\left.+Y_{\text{hi}}^4 \left(Y_{\text{ve}}^2
   \left(Y_{\text{vi}} \left((Y (Y ((Y-2) Y+5)-2)+1) Y_{\text{vi}}-((Y-5) Y+1)
   ((Y-1) Y+1)\right)
\right.\right.\right.\right.\nn\\   
   & \left.\left.\left.\left.+Y (Y ((Y-2) Y+5)-2)+1\right)+Y_{\text{ve}}
   \left(Y_{\text{vi}} \left(-((Y-5) Y+1) ((Y-1) Y+1) Y_{\text{vi}}
\right.\right.\right.\right.\right.\nn
\\   
&\left.\left.\left. \left.\left.
   +5 Y (Y
   ((Y-2) Y+5)-2)+5\right)
   -((Y-5) Y+1) ((Y-1) Y+1)\right)+Y_{\text{vi}}
   \left((Y (Y ((Y-2) Y+5)-2)+1) Y_{\text{vi}}
\right.\right.\right.\right.\nn
\\   
   &\left.\left.  \left.\left.    
   -((Y-5) Y+1) ((Y-1) Y+1)\right)
   +Y
   (Y ((Y-2) Y+5)-2)+1\right)
+Y^2 \left(Y_{\text{ve}}^2+Y_{\text{ve}}+1\right)
   \left(\left(Y_{\text{vi}}-1\right) Y_{\text{vi}}+1\right)\right)
\right.\nn\\   
  &\left.
   +3
   Y_{\text{he}}^2 \left(Y^2 Y_{\text{hi}}^6
   \left(Y_{\text{ve}}^2+Y_{\text{ve}}+1\right)
   \left(\left(Y_{\text{vi}}-1\right) Y_{\text{vi}}+1\right)+2 Y
   \left(Y^2+Y+1\right) Y_{\text{hi}}^4 \left(Y_{\text{ve}} \left(Y_{\text{vi}}
   \left(Y_{\text{ve}}+Y_{\text{vi}}+2\right)+1\right)+Y_{\text{vi}}\right)
   \right. \right.\nn\\   
  &\left.\left.
   +Y_{
   \text{hi}}^2 \left(Y_{\text{ve}}^2 \left(Y_{\text{vi}} \left((Y (Y ((Y-2)
   Y+5)-2)+1) Y_{\text{vi}}-((Y-5) Y+1) ((Y-1) Y+1)\right)
\right.\right.\right.\right.\nn\\   
  &\left.  \left.  \left.  \left.   
   +Y (Y ((Y-2)
   Y+5)-2)+1\right) 
   +Y_{\text{ve}} \left(Y_{\text{vi}} \left(-((Y-\!5) Y+1) ((Y-\!1)
   Y+\!1) Y_{\text{vi}}
 +5 Y (Y ((Y-\!2) Y+\!5)\!-\!2)\!+\!5\right)
\right.\right.\right.\right.\nn\\   
  &\left.\left.\left. \left.
 -((Y-5) Y+1) ((Y-1)
   Y+1)\right)
   +Y_{\text{vi}} \left((Y (Y ((Y-2) Y+5)-2)+1) Y_{\text{vi}}-((Y-5)
   Y+1) ((Y-1) Y+1)\right)
   \right.\right.\right.\nn\\   
  &\left.\left.  \left.
   +Y (Y ((Y-2) Y+5)-2)+1\right)
   +2 Y ((Y-1) Y+1)
   Y_{\text{ve}} \left(\left(Y_{\text{vi}}-1\right)
   Y_{\text{vi}}+1\right)\right)+\left(\left(Y_{\text{ve}}-1\right)
   Y_{\text{ve}}+1\right) 
 \right.\nn\\   
 &     \left.  
   \left(3 Y^2 Y_{\text{hi}}^4
   \left(Y_{\text{vi}}^2+Y_{\text{vi}}+1\right)
     +6 Y ((Y-1) Y+1) Y_{\text{hi}}^2
   Y_{\text{vi}}+((Y-1) Y+1)^2 \left(\left(Y_{\text{vi}}-1\right)
   Y_{\text{vi}}+1\right)\right)\right)\,. \label{eq:looppol_torus}
\end{align}
Expanding this expression, we find monomials such as $Y^{3}$ or $Y_{\text{hi}}^{2}Y_{\text{he}}^{2}$ corresponding to non-contractible cycles, and other terms such as $YY_{\text{ve}}Y_{\text{he}}^{2}$ or $YY_{\text{vi}}Y_{\text{hi}}^{2}$ corresponding to contractible loops drawn on the torus.

We may now apply the formula \eqref{eq_Ising_zeroes} to compute the geometric ansatz for Ising zeros, taking into account that the vertical inner edges are concave, thus with negative dihedral angle $\theta_{\text{vi}}=-2\pi/3$, and that we use \eqref{eqn:circlepattern} for the angles of all the quadrilateral faces:
\be
Y^{(0)}
=
\frac{2 \sqrt{r^2+r R+R^2}-\sqrt{3}\, (R+r)}{R-r}
\,,
\ee
\be
Y_{\text{vi}}^{(0)}
=
e^{-\frac{i \pi }{3}}\,\frac{ \left(\sqrt{h^2+3
   r^2}-\sqrt{3}\, r\right)}{h}
\,,\qquad
Y_{\text{hi}}^{(0)}
=
e^{\frac{i \pi }{4}}\,\frac{ \sqrt{\left(-h+\sqrt{h^2+3
   r^2}\right) \left(-r-2 R+2 \sqrt{r^2+r
   R+R^2}\right)}}{\sqrt{3}\, r}
\,,\nn
\ee
\be
Y_{\text{ve}}^{(0)}
=
e^{\frac{i \pi }{3}}\frac{ \left(\sqrt{h^2+3
   R^2}-\sqrt{3} R\right)}{h}
\,,\qquad
Y_{\text{he}}^{(0)}
=
e^{\frac{i \pi }{4}}\frac{ \sqrt{\left(-h+\sqrt{h^2+3 R^2}\right)\left(2 r+R+2 \sqrt{r^2+r
   R+R^2}\right) }}{\sqrt{3}\, R}
\,.\nn
\ee
%

A typical configuration is given by the edge lengths $(r,R,h)=(1,2,1)$, for which  the absolute value of the loop polynomial\footnotemark{} is around  $5.4\times 10^{-2}$, or more precisely
\be
P_{\text{torus}}\big{[}\{Y_{\ell}^{(0)}\}\big{]}
\sim
0.0437193 - 0.0318252 i
\,,\qquad
\Big{|}
P_{\text{torus}}\big{[}\{Y_{\ell}^{(0)}\}\big{]}
\Big{|}
\sim
0.054076
\,,\nn
\ee
which is a very poor result for the low number of vertices of this configuration (12 vertices) compared with the case of the sphere (see fig.\ref{fig_precission}).
\footnotetext{
Let us emphasize that we computed here the value of the loop polynomial, which is  the full partition function divided by $2^{12}=4096$.
}
In fact, we can move around the values of the edge lengths and see that the value of loop polynomial on the geometric ansatz $Y_{\ell}^{(0)}$ given above  varies and does not vanish.
\begin{figure}[ht!]
\centering
\vspace*{3mm}
\subfloat[Plot as $r$ varies from 0 to 2 for $R=2$ and $h=1$.]{
\begin{overpic}[abs,unit=1mm,width=5cm
]{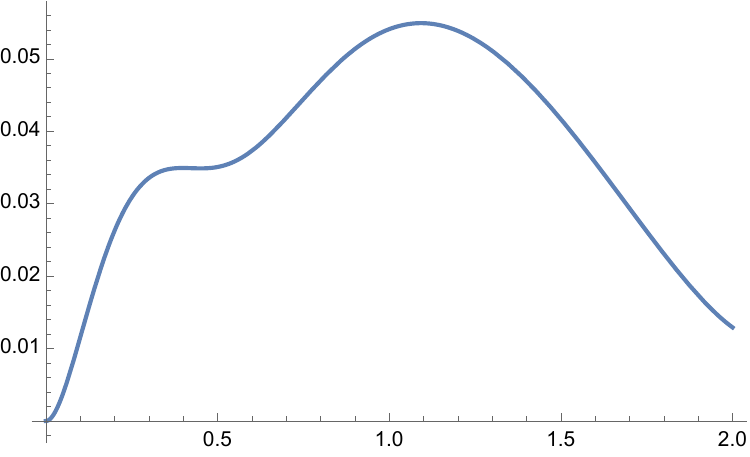}
\put(-2,33){{\scriptsize $\Big{|}
P_{\text{torus}}\big{[}\{Y_{\ell}^{(0)}\}\big{]}
\Big{|}$}}
\put(52,2){$r$}
\end{overpic}
}
\hspace{8mm}
\subfloat[Plot as $R$ varies from 1 to 10 for $r=1$ and $h=1$.]{
\begin{overpic}[abs,unit=1mm,width=5cm
]{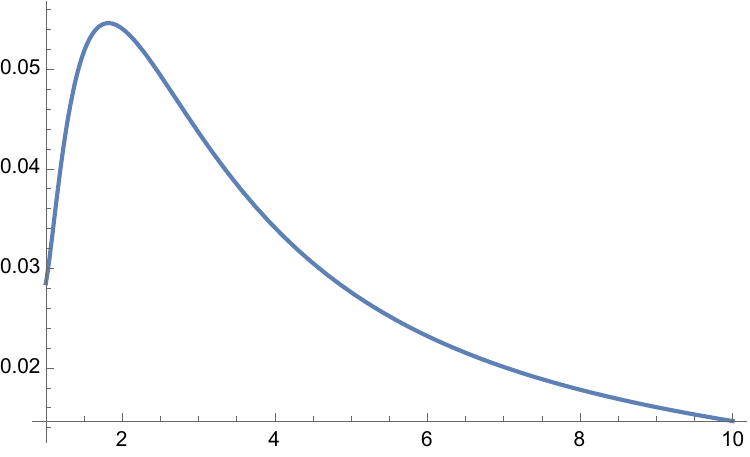}
\put(-2,33){{\scriptsize $\Big{|}
P_{\text{torus}}\big{[}\{Y_{\ell}^{(0)}\}\big{]}
\Big{|}$}}
\put(52,2){$R$}
\end{overpic}}
\hspace{8mm}
\subfloat[Plot as $h$ varies from 0 to 10 for $r=1$ and $R=2$.]{
\begin{overpic}[abs,unit=1mm,width=5cm
]{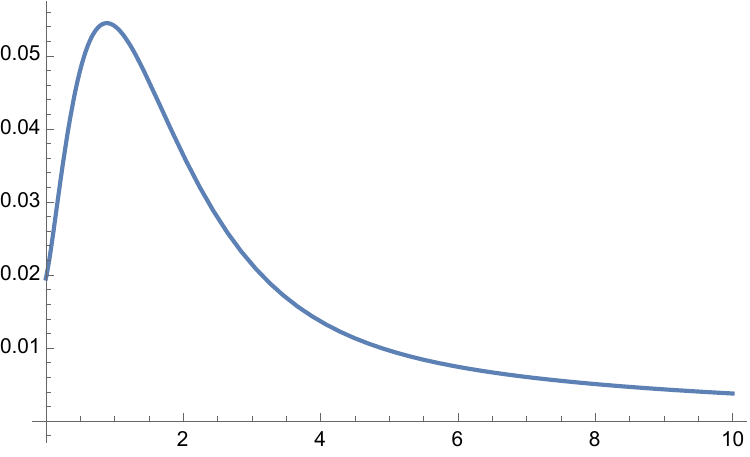}
\put(-2,33){{\scriptsize $\Big{|}
P_{\text{torus}}\big{[}\{Y_{\ell}^{(0)}\}\big{]}
\Big{|}$}}
\put(51.5,2){$h$}
\end{overpic}
}
\caption{
Plots of the absolute value of the loop polynomial $P_{\text{torus}}$ of the prismatic torus evaluated on the geometric ansatz $Y_{\ell}^{(0)}$ as the edge lengths $r,R,h$ vary.
\label{plot_torus}
}
\end{figure}

\begin{figure}[ht!]
\centering
\vspace*{2mm}
\subfloat[Plot as $R$ varies from 100 to 2000 for $r=1$ and $h=1$.]{
\begin{overpic}[abs,unit=1mm,width=6.5cm
]{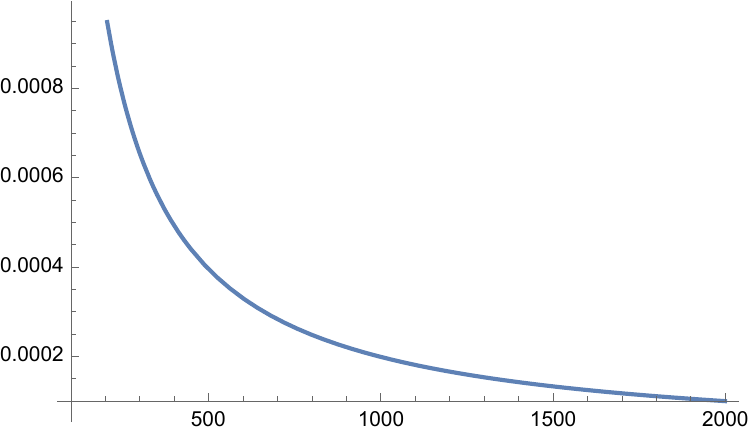}
\put(0,41){{\scriptsize $\Big{|}
P_{\text{torus}}\big{[}\{Y_{\ell}^{(0)}\}\big{]}
\Big{|}$}}
\put(67,2){$R$}
\end{overpic}
}
\hspace{20mm}
\subfloat[Plot as $h$ varies from 10 to 100 for $r=1$ and $R=2$.]{
\begin{overpic}[abs,unit=1mm,width=6.5cm
]{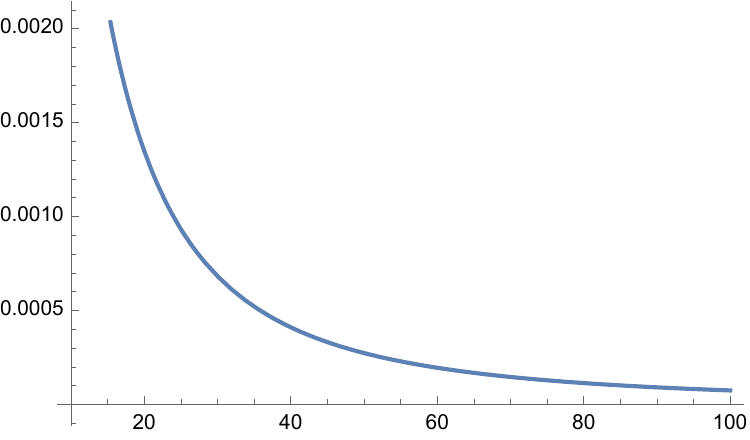}
\put(-2,41){{\scriptsize $\Big{|}
P_{\text{torus}}\big{[}\{Y_{\ell}^{(0)}\}\big{]}
\Big{|}$}}
\put(67,2){$h$}
\end{overpic}
}
\caption{
Plots of the absolute value of the loop polynomial $P_{\text{torus}}$ of the prismatic torus evaluated on the geometric ansatz $Y_{\ell}^{(0)}$ in the asymptotics $R\rightarrow \infty$ (on the left side) and $h\rightarrow \infty$ (on the right side).
\label{plot_torus-asympt}
}
\end{figure}
On fig.\ref{plot_torus}, we see that the loop polynomial does not vanish for the geometric couplings, which is clearly different from the case with spherical topology. Nevertheless, on fig.\ref{plot_torus-asympt}, we notice that it does vanish in the asymptotic limit as $r\rightarrow 0$ or as $R\rightarrow\infty$ or as $h\rightarrow\infty$. Precise numerics are given in tables \ref{table:asympt-torus}.
We do not have an analytical understanding for these limits, but one could argue that in the limits $r\rightarrow 0$ or $R\rightarrow\infty$, the hole of the donut becomes smaller and smaller, even though this mechanism does not seem to apply to the limit of large height  $h\rightarrow\infty$. It would definitely be enlightening to understand why the geometrical formula for the spherical topology applies to these asymptotic limits of the toroidal topology and whether this feature should be expected to extend to higher genus topologies.
\begin{table}[ht]
\centering
\bgroup
\def\arraystretch{1.3}
\begin{tabular}{ || c | c |}
\hline
$r$  & $\big{|}P_{\text{torus}}(r,R=2,h=1)\big{|}$  \\ \hline 
 $10^{-1}$ & $1.15107\times 10^{-2}$  \\  \hline 
  $10^{-2}$ & $1.87652 \times 10^{-4}$ \\ \hline
  $10^{-3}$ & $1.96812 \times 10^{-6}$ \\ \hline
  $10^{-4}$ & $1.97749 \times 10^{-8}$ \\ \hline
\end{tabular}
\hspace*{15mm}
\begin{tabular}{ || c | c |}
\hline
$R$  & $\big{|}P_{\text{torus}}(r=1,R,h=1)\big{|}$  \\ \hline
 $10^{2}$ & $1.90415\times 10^{-3}$  \\  \hline 
  $10^{3}$ & $1.98433 \times 10^{-4}$ \\ \hline
  $10^{4}$ & $1.99279 \times 10^{-5}$ \\ \hline
  $10^{5}$ & $1.99355 \times 10^{-6}$ \\ \hline
\end{tabular}
\hspace*{15mm}
\begin{tabular}{| | c | c |}
\hline
$h$  & $\big{|}P_{\text{torus}}(r=1,R=2,h)\big{|}$  \\ \hline
 $10^{2}$ & $7.3781\times 10^{-5}$  \\  \hline 
  $10^{3}$ & $7.91509 \times 10^{-7}$ \\ \hline
  $10^{4}$ & $7.97082 \times 10^{-9}$ \\ \hline
  $10^{5}$ & $7.97643 \times 10^{-11}$ \\ \hline
\end{tabular}
\egroup
\caption{Numerics for the asymptotics of the evaluation of the loop polynomial on the geometric couplings as $r\rightarrow 0$ on the left, $R\rightarrow \infty$ in the center, $h\rightarrow \infty$ on the right.
\label{table:asympt-torus}}
\end{table}


We have also considered another toroidal configuration: the wedgy donut is made of three wedges glued together by their triangular ends into a ring, as drawn on fig.\ref{fig:isosceles_donut} and fig.\ref{fig:equilateral_donut}. We focussed on two specific configurations, with either an isosceles corner triangle or an equilateral corner triangle. Although the values of the dihedral and planar angles are more intricate and the loop polynomial more complicated, these two configurations both behave similarly to the prismatic torus, with numerics of the same order of magnitude.

At the end of the day, even if the toroidal configurations which we  analyzed can not be considered as generic or general, they are enough to conclude that the topology has a non-trivial impact on the validity of the geometric formula for the Ising zeros and that an appropriate extension is required to handle non-planar graphs and toroidal triangulations. Such a generalization to non-trivial surface topologies requires a deeper exploration of the duality formulas between the 2d  Ising model and 3d quantum gravity, which is clearly beyond the scope of the present study and which we postpone to future investigation.

%

\section*{Conclusion \& Outlook}

The holographic duality discovered between 3d quantum gravity and the 2d inhomogeneous Ising model, implemented by an exact equality holding at the discrete level and away from criticality,  led to a geometric formula for the critical couplings \cite{Bonzom:2015ova,Bonzom:2019dpg,Bonzom:2024zka}.
More precisely, it is a formula for the zeros of the Ising partition function on finite planar graphs in terms of the embedding of triangulations, dual to the graph, in the flat 3d space. This formula was derived by an asymptotic analysis of the saddle point of the quantum gravity path integral in its semi-classical regime. Although proven explicitly in the case of the tetrahedral graph in \cite{Bonzom:2019dpg}, the derivation of the general case presented in \cite{Bonzom:2024zka} relies on asymptotical approximation, whose precision is not entirely under control. The geometric formula for Ising zeros therefore remains a mathematical conjecture.

The purpose of the present study was twofold: on the one hand, provide explicit examples with simple geometrical configurations, on top of the already-studied tetrahedron, and on the other hand, provide a thorough numerical check on random polyhedra with many faces, thus dual to planar graphs with many nodes.

First, working out explicit examples is both a pedagogical effort to show how to apply the formula for planar graphs, in terms of the geometry of their dual triangulations or circle patterns, and an analytical effort to show explicitly the validity of the formula for other graphs than the tetrahedral graph.

Second, we performed a thorough numerical checks with random polyhedra, with both convex and no-convex folds, for up to 30 faces, validating the geometric formula for Ising zeros in terms of the dihedral and planar angles of the polyhedra. Our method was to generate random polyhedra with triangular faces using locally-rescaled Delaunay triangulations for random sprinkling of the unit 2-sphere.
This provides strong support for the proposed Ising zeros formula.

A side-product of our analysis is a clarification of the sign prescription for the dihedral angles encoding the embedding (and more precisely, the extrinsic curvature) of the 2d triangulation in the 3d flat Euclidean space, and how that sign reflect the convexity or non-convexity of the local folds between the faces.

\smallskip

Now, with such numerical validation, one still needs to find a mathematically-clean proof of the formula. Moreover, it would also be enlightening to generalize that Ising zero geometric formula to higher topologies. Indeed, we also showed here that the present formula does not work for a toroidal topology. This result underlines the importance of the assumption of planar graphs and spherical topology in the original work in \cite{Bonzom:2015ova,Bonzom:2024zka}, and highlights the necessity of the extension of the QG${}_{3d}$$\,\leftrightarrow\,$Ising${}_{2d}$ duality to  higher genus surface topologies, where non-trivial transport can develop around non-contractible cycles. This requires understanding how such non-trivial holonomies affect the Ising partition function.
Indeed, earlier mathematical work shows that the Ising model on higher genus corresponds to spin network evaluations with specific holonomy observable insertions \cite{CostantinoHDR}. This claim remains to be translated in terms of 3d quantum gravity correlation functions.
This line of research promises an interesting interplay between quantum gravity and statistical physics, on critical Ising couplings for periodic lattices, topology-protected band structures and edge modes propagating on 2d surfaces in quantum gravity.

\section*{Acknowledgement}
This work is supported by the Basque Government Grant IT1628-22, and by the Grant PID2021-123226NBI00 (funded by MCIN/AEI/10.13039/501100011033 and by ``ERDF A way of making Europe'').

%
%




\providecommand{\href}[2]{#2}\begingroup\raggedright\endgroup

\end{document}